\pdfoutput=1
\documentclass[authoryear,preprint,review,11pt]{elsarticle}

\usepackage{amssymb}
\usepackage{amsmath}
\usepackage{mathtools}
\usepackage{IEEEtrantools}
\usepackage[table,xcdraw,svgnames,dvipsnames]{xcolor}
\usepackage{url,lineno,microtype,subcaption}
\usepackage[onehalfspacing]{setspace}
\usepackage{bm}
\usepackage{scalerel}
\usepackage{mathrsfs}
\usepackage{relsize}
\usepackage{textcomp}
\usepackage{gensymb}
\usepackage{graphicx}
\usepackage[colorlinks]{hyperref}
\usepackage{upgreek}
\usepackage{textgreek}
\usepackage[T1]{fontenc}
\usepackage{parskip}
\usepackage{media9}
\usepackage{listings}
\usepackage{float}
\usepackage{algorithm}
\usepackage[noend]{algpseudocode}
\usepackage{textcomp}
\usepackage{comment}

\definecolor{cerulean}{rgb}{0.0, 0.48, 0.65}
\AtBeginDocument{%
  \hypersetup{
    citecolor=cerulean,
    linkcolor=Red,   
    urlcolor=Magenta}}
    
\graphicspath{ {./figures/} }

\definecolor{mygreen}{RGB}{28,172,0} 
\definecolor{mylilas}{RGB}{170,55,241}
\journal{\,}

\begin{document}
\begin{frontmatter}

\title{Mode Decomposition-based Time-varying Phase Synchronization for fMRI}

\author[A1]{Hamed Honari}
\author[A2]{Martin A. Lindquist\corref{cor1}}
\cortext[cor1]{Corresponding author}
\ead{mlindqui@jhsph.edu}
\address[A1]{Department of Electrical and Computer Engineering, Johns Hopkins University, USA}
\address[A2]{Department of Biostatistics, Johns Hopkins University, USA}

\begin{abstract}
Recently there has been significant interest in measuring time-varying functional connectivity (TVC) between different brain regions using resting-state functional magnetic resonance imaging (rs-fMRI) data. One way to assess the relationship between signals from different brain regions is to measure their phase synchronization (PS) across time.  However, this requires the \textit{a priori} choice of type and cut-off frequencies for the bandpass filter needed to perform the analysis.  Here we explore alternative approaches based on the use of various mode decomposition (MD) techniques that circumvent this issue. These techniques allow for the data driven decomposition of signals jointly into narrow-band components at different frequencies, thus fulfilling the requirements needed to measure PS. We explore several variants of MD, including empirical mode decomposition (EMD), bivariate EMD (BEMD), noise-assisted multivariate EMD (na-MEMD), and introduce the use of multivariate variational mode decomposition (MVMD) in the context of estimating time-varying PS. We contrast the approaches using a series of simulations and application to rs-fMRI data.  Our results show that MVMD outperforms other evaluated MD approaches, and further suggests that this approach can be used as a tool to reliably investigate time-varying PS in rs-fMRI data.

\end{abstract}


\end{frontmatter}

\section{Introduction}
In recent years, there has been a great deal of interest in studying time-varying functional connectivity (TVC) in the brain during the course of a single resting-state functional magnetic resonance imaging (rs-fMRI) run; see, for example, \cite{lurie2019questions} for an overview. A number of different analytic approaches have been suggested for measuring TVC, including the correlation-based sliding window approach \cite{tagliazucchi2012dynamic, chang2010time, hutchison2013dynamic, hutchison2013resting}, change point analysis  \cite{cribben2012dynamic, cribben2013detecting, xu2015dynamic}, point process analysis \cite{tagliazucchi2011spontaneous}, co-activation patterns (CAPs) \cite{liu2013time, karahanouglu2015transient}, time series models \cite{lindquist2014evaluating}, time-frequency analysis \cite{chang2010time}, and variants of hidden Markov models (HMMs)  \cite{eavani2013unsupervised, vidaurre2017brain, shappell2019improved, bolton2018interactions}.

Recently, phase synchronization (PS) methods have gained increased popularity in fMRI analysis as a means of measuring the level of synchrony between the phase of time series from different regions of interest (ROIs) \cite{glerean2012functional, pedersen2017spontaneous, rebollo2018stomach, pedersen2018relationship,choe2021phase, honari2021evaluating, choe2021methodological} .  The term {\em synchrony} refers to the coordination in the state of two or more systems that can be attributed to their interaction or coupling \cite{rosenblum1996phase}. In these methods the phase of a set of time series from different ROIs are computed at each time point through the application of the Hilbert transform and used to evaluate the phase difference between the time series. Two time series in synchronization will maintain a constant phase difference, and time-varying PS methods seek to investigate how the phase difference between time series from different regions vary during the course of a rs-fMRI run. Throughout we will differentiate between methods that combine a PS metric with a sliding window approach, referred to as windowed phase synchronization (WPS), with those that directly measure PS at each time point, referred to as instantaneous phase synchronization (IPS) \cite{honari2021evaluating}. 

To date, a number of studies have applied PS methods to fMRI data.  An early application was \cite{laird2002characterizing}, who used IPS to analyze task-activated fMRI data. Later \cite{niazy2011spectral} studied the spectral characteristics of resting state networks (RSNs), highlighting the importance of considering IPS between RSNs at different frequencies, and \cite{glerean2012functional} proposed the use of IPS as a measure of TVC.  \cite{pedersen2018relationship} examined the relationship between IPS and correlation-based sliding window (CSW) techniques and observed a strong association between the two methods when using absolute values of CSW. Finally, \cite{honari2021evaluating} critically evaluated a number of different WPS and IPS methods for evaluating PS, introducing several new WPS method and a new method within the IPS framework denoted the cosine of the relative phase (CRP). Their results indicated that using CRP within an IPS framework outperformed other tested methods.

A benefit of using the IPS approach is that it offers single time-point resolution of time-resolved fMRI connectivity and does not require choosing an arbitrary window as is the case when using, for example, CSW.  However, a shortcoming involves the need to bandpass filter the data prior to analysis in order to accurately estimate the instantaneous phases. This requirement stems from the fact that in order for the instantaneous phases obtained using the Hilbert transform to be physically meaningful, the signal must be sufficiently narrow-bandpassed to approximate a monocomponent signal. To circumvent the need to \textit{a priori} choose the type and cut-off frequencies for the bandpass filter, researchers are increasingly exploring the use of various mode decomposition (MD) techniques that allow for the data driven decomposition of signals into narrow-band components centered at different frequencies.
As an example, the Hilbert-Huang Transform (HHT) \cite{huang1998empirical} combines a MD technique, Empirical Mode Decomposition (EMD), with the Hilbert transform.  The EMD step decomposes the signal into a finite set of time-varying oscillatory functions referred to as intrinsic mode functions (IMFs).  
The second step of the HHT involves the application of the Hilbert transform to each IMF generated using EMD to obtain their time-varying measures of instantaneous phase. The HHT provides a data-driven tool for time-frequency analysis, that is able to handle signals that are inherently non-linear and non-stationary. In this paper we similarly focus on using EMD and other related MD techniques as alternatives to bandpass filtering the data prior to applying the Hilbert transform and evaluating PS.

There exists multiple types of MD techniques that can be used to decompose the data into IMFs.  The simplest version is standard EMD which can be applied to a single univariate signal. This procedure is performed through a recursive algorithm that uses the local extrema in the signal to estimate its upper and lower envelopes.  These envelops are then used to obtain IMFs using the so-called sifting process.  This involves averaging the envelopes to compute the local mean, thus providing a low-frequency estimate of the signal. Removing the local mean from the signal allows one to recover a high-frequency oscillatory mode (i.e., an IMF). The process is repeated until all oscillatory modes in the signal have been obtained. It has been demonstrated \cite{flandrin2004empirical, wu2004study, rehman2010multivariate} that IMFs obtained using standard EMD provides frequency responses similar to those obtained using a dyadic filter bank (i.e., a filter bank dividing a signal into a collection of successively band-limited components in which while descending the frequency scale, successive frequency bands have half the width of their predecessors \cite{vouras2005principal}. In other words, EMD behaves as a filter bank of overlapping bandpass filters and the frequency content of the IMFs obtained decreases from one IMF to the following one. 

EMD is designed to be performed on univariate signals. In order to perform a bivariate or multivariate analysis, the decomposition needs to be performed simultaneously and jointly for several reasons. First, the computation of the local mean, which depends on finding local extrema for multivariate signals, is not straightforward.  This is because the concept of local extrema is not well defined for multivariate signals and the notion of the oscillatory modes defining an IMF is convoluted \cite{rehman2010multivariate, mandic2009complex}. Second, to allow for mode alignment across signals and alleviate frequency mismatch between the decomposed IMFs, they need to be matched both in number and frequency \cite{looney2009multiscale}.  

Bivariate EMD (BEMD) \cite{tanaka2007complex} combines the two signals being analyzed into a single complex signal and thereafter computes its envelopes. If analyzing more than two signals this approach can be applied in a pair-wise manner.  However, in this setting the alignment of the bivariate IMFs among all pairs is not guaranteed globally.  This necessitated the development of multivariate EMD (MEMD) which ensures global mode alignment across all IMFs \cite{rehman2010multivariate}. This is achieved by forming multiple real-valued projections of the signal, and using them to obtain  multi-dimensional envelopes via interpolation.  

While MEMD addresses mode alignment and guarantees that the same number of IMFs are extracted from each multivariate input signal, it is susceptible to mode mixing (i.e., the mixture of frequency content from one IMF to another). In addition, the manner in which IMFs are chosen can be driven by noise.  This led to the introduction of noise-assisted multivariate EMD (na-MEMD) \cite{yeh2010complementary, colominas2012noise, zhang2017noise} where additional input signals generated from Gaussian white noise (WN) are introduced.  The structure of the frequency response that MEMD imposes on Gaussian WN is similar to that of a dyadic filter bank \cite{vouras2005principal}). 
Hence, MEMD behaves as a bank of bandpass filters and the frequency content of the obtained IMFs decrease in subsequent IMFs.  Thus, when Gaussian WN is added to the input signal this allows the intrinsic oscillations to be filtered adaptively to their appropriate scales. 

Another noise assisted variant of MD includes ensemble EMD (EEMD) which leverages the dyadic filter bank property by adding Gaussian WN directly to the signal \cite{wu2009ensemble}.  This procedure is repeatedly performed, and for each iteration a different realization of WN is used. The output IMFs are obtained by averaging the corresponding IMFs from the whole ensemble. However, determining optimal values for the noise level and number of ensembles is not trivial.
 
Further, adding too much noise directly to the signal can degrade results. This approach is outside the scope of this paper as previous work has shown that na-MEMD outperforms EEMD \cite{zhang2017noise}. 

It should be noted that the EMD variants discussed earlier are only partially able to address the limitations on sensitivity to noise and mode mixing. This is true because EMD is based on the use of signal extrema, and in general the extrema of a sampled signal may differ from its continuous-time version.  A consequence of that is the local mean may introduce artefacts associated with how sampling is performed \cite{rilling2009sampling}. 
In addition, EMD is a recursive process whose results depend on various 
algorithmic choices, including which interpolation scheme and stopping criteria is used \cite{ur2019multivariate}. Given these concerns, \cite{dragomiretskiy2013variational} introduced a non-recursive Variational Mode Decomposition (VMD) technique where the modes are extracted concurrently.  Here an optimization problem is formulated where the cost function to be minimized is the sum of bandwidths of all signal modes subject to the constraint that the modes reconstruct the input signal.  \cite{ur2019multivariate} introduced a generic extension of VMD for multivariate data, referred to as Multivariate Variational Mode Decomposition (MVMD), where the cost function is extended to minimize the sum of bandwidths of all signal modes across all multivariate input signals.  

EMD approaches have already been successfully applied to rs-fMRI data. For example, \cite{honari2021evaluating} compared PS analysis on rs-fMRI data using data extracted via the Hilbert transform and BEMD.  Further, \cite{zhou2020prediction} used MEMD to analyze TVC in rs-fMRI for use in prediction and classification of sleep quality.  In this work, we therefore primarily focus on the use of na-MEMD and MVMD for evaluating TVC using rs-fMRI within the PS framework. 

It is important to note that there exist several alternative approaches towards decomposing signal into frequency components. For example, the Fourier transform is well-suited when the data is linear and periodic or stationary. However, the Fourier basis functions are global, and thus cannot handle local non-linearity without significant spreading across frequencies.  This is particularly true when the waveforms deviate significantly from a sinusoidal form.  Another example, the wavelet transform, is basically an adjustable window Fourier transform, which can capture localised information in the time-frequency domain due to its multi-scale property. It therefore provides a better approach than the Fourier transform for non-stationary signals.  However, this approach has a non-adaptive nature and once the mother wavelet is chosen, it cannot be changed during the analysis. Further, it is not suitable for the analysis of non-linear data.   In contrast, MD techniques are data driven and adaptive in terms of decomposing the signal into modes. Their ability to handle non-linear, non-stationary time series make them particularly suitable for the analysis of rs-fMRI time series data \cite{guan2020profiles, allen2014tracking}.

The paper is organized as follows. We begin by introducing standard univariate EMD and describing the algorithm used to decompose an input signal into IMFs. Next, we discuss variants of EMD that can be used for multivariate decomposition of multiple signals jointly into IMFs rather than decomposing each signal independently.  These include BEMD, MEMD, and na-MEMD. Finally, MVMD is introduced as an alternative decomposition approach that mitigates issues related to mode mixing and sensitivity to noise often observed in EMD approaches. The paper concludes by comparing the methods using a series of simulations, and applying MVMD-based phase synchronization to rs-fMRI data.

\section{Methods}

\subsection{Empirical Mode Decomposition Techniques}

To obtain the instantaneous phase \cite{boccaletti2018synchronization} of a real signal $x(t)$ one must first construct an analytic signal:
\begin{equation}
    z(t) = x(t) + j \mathcal{H}\{x(t)\}
\end{equation}
where $j = \sqrt{-1}$ and $\mathcal{H}$ represents the Hilbert Transform. This signal can be re-expressed as follows:
\begin{equation}
    z(t) = \mathfrak{A}(t) \exp\big(j {\phi}(t)\big)
\end{equation}
where $\mathfrak{A}(t)$ represents the envelope and ${\phi}(t)$ the instantaneous phase. 

An important consideration when computing instantaneous phase is the need for $x(t)$ to satisfy Bedrosian's Product Theorem \cite{bedrosian1963product}, which states that a band-limited signal can be decomposed into the product of envelope and phase when their spectra are disjoint.  A necessary condition for this to hold is that $x(t)$ is first narrow-banded by applying a bandpass filter. Choosing the type and bandwidth for the bandpass filter required to perform the analysis is one of the greater challenges when using these types of approaches.  This drives our motivation to investigate alternative approaches that do not depend on the restrictive conditions imposed by Bedrosian's theorem.  

One such alternative is the Hilbert Huang transform, which is EMD followed by the Hilbert transform. Here EMD is used in place of bandpass filtering. 
EMD offers a data-driven way to decompose the signal of interest into a series of IMFs that correspond to different frequency bands. An IMF is a function where the number of extrema and zero-crossings differ at most by one, and the mean value of the upper (defined by the local maxima) and lower envelopes (defined by the local minima) is equal to zero. The IMFs are zero-mean amplitude-frequency modulated signals, designed to ensure that Bedrosian's theorem is respected and that subsequent application of the Hilbert transform will result in physically meaningful instantaneous phases of the input signal \cite{ur2011filter, huang2014hilbert}. Together, these IMFs form a set of approximately orthogonal basis that can be used to reconstruct the input signal. The first IMF consists of the largest frequency oscillation present in the signal, and each subsequent IMF consists of increasingly smaller frequency oscillations than those previously extracted. When using EMD there are no \textit{a priori} assumptions required for the data, making it an ideal approach for the analysis of nonlinear and non-stationary processes \cite{park2011time} such as those observed in rs-fMRI. 

The practical implementation of EMD is described in Algorithm \ref{alg:emd}, where Steps 2-6 are referred to as the \textit{sifting process}.  The first step is to identify all the local extrema present in the input signal $x(t)$. All local maxima are connected using a cubic spline which create an upper envelope for the signal. Similarly, all local minima are used to create a lower envelope. 
The two envelopes create an upper and lower bound for the data. The mean $m(t)$ is computed by taking their average. The difference between the data and mean, $d(t) = x(t) - m(t)$, is considered a candidate IMF.  This process is repeated until a stoppage criterion is fulfilled, and the current value of $d(t)$ is designated as an IMF. Once, the IMF is set it is subtracted from the signal and the sifting process is performed on the new signal. This process is repeated, each time removing the previously defined set of IMFs from $x(t)$, until the desired number of IMFs have been obtained.   The stoppage criterion used in EMD determines the number of sifting steps that are performed when deriving IMFs.  

There are a number of commonly used stoppage criterion,and in this paper we use the threshold method \cite{rilling2003empirical}. This criterion is based on imposing two thresholds $\theta_1$ and $\theta_2$ aimed at ensuring globally small fluctuations in the mean while taking locally large excursions into consideration \cite{rilling2003empirical}. Let us define the amplitude as $a(t):= (e_{max}(t) - e_{min}(t))/2$, 
where $e_{max}(t)$ and $e_{min}(t)$ are the upper and lower envelope, respectively, and the evaluation function as $\sigma(t):=|m(t)/a(t)|$.  The sifting process is repeated until $\sigma(t) < \theta_1$ for some fraction $(1-\alpha)$ of the total number of iterations 
and $\sigma(t)<\theta_2$ for the remaining fraction.  The recommended values for these thresholds proposed by \cite{rilling2003empirical}, which we use in this paper, are $\alpha \approx 0.05$, $\theta_1 \approx 0.05$, and $\theta_2 \approx 10 \theta_1$.

\begin{algorithm}
\caption{Standard Empirical Mode Decomposition}\label{alg:emd}
\begin{algorithmic}[1]

\Procedure{EMD($x(t)$, stoppage criteria, number of IMFs)}{} \Comment{$x(t)$ is the input signal }
\State $\Tilde{x}(t) \gets x(t)$
\State Extract all local extrema of $\Tilde{x}(t)$
\State $e_{max}(t),e_{min}(t)\gets$ form upper and lower envelopes from all local maxima and minima, respectively, using cubic splines
\State $m(t)\gets \frac{e_{min}(t)+e_{max}(t)}{2}$ 
\State $d(t) \gets \Tilde{x}(t)-m(t)$ 
\State $\Tilde{x}(t) \gets d(t)$ go to line 3; \textbf{repeat} until $d(t)$ becomes an IMF \Comment{Stoppage Criteria}
\State \textbf{repeat} lines 3:7 until desired number of IMFs obtained 
\EndProcedure
\end{algorithmic}
\end{algorithm}

\subsubsection{Bivariate Empirical Mode Decomposition}

Often it is of interest to extract IMFs from a collection of signals.
A simple approach is to apply the standard EMD algorithm separately to each signal and extract a set of IMFs from each. However, this does not guarantee that the frequencies of the extracted IMFs match across signals, nor will the repeated application of the EMD algorithm necessarily produce the same number of IMFs for all signals. This is particularly problematic within the PS framework, where one seeks to compare signals within the same frequency band.  To circumvent this issue and allow for the decomposition of a bivariate or multivariate signal, various extensions of EMD have been proposed that perform the decomposition jointly \cite{ur2011filter}.  
To decompose bivariate signals, EMD should be performed so that the decomposition and sifting process is applied on the envelope of the bivariate signal.  This ensures that the number and frequency of the decomposed IMFs corresponding to each signal match. \cite{rilling2007bivariate} proposed an extension of EMD, referred to as BEMD, for handling signal pairs.  Here the bivariate signal is treated as a complex-valued signal $x(t)$ to simplify representation. In general, given a set of directions in the complex plane, BEMD follows the same algorithm as EMD except that the sifting process is performed jointly on the signal pair.  Thus, for bivariate signals the looping of the algorithm is performed in the directions $\pi$ and $2\pi$. In each direction all extrema are extracted and the set of points interpolated using cubic splines to obtain the envelopes. The envelopes corresponding to these directions are averaged to obtain the local mean. The practical implementation of BEMD is described in Algorithm \ref{alg:bemd}.

\begin{algorithm}
\caption{Bivariate Empirical Mode Decomposition}\label{alg:bemd}
\begin{algorithmic}[1]
\Procedure{BEMD($x(t)$,stoppage criteria, number of IMFs)}{}\Comment{$x(t)$ is the input signal}
\State $\Tilde{x}(t) \gets x(t)$
\State \textbf{for} $1\leq k \leq N$ \textbf{do}
\State \hspace{0.5cm} Project the complex-valued signal $x(t)$ on direction $\varphi_k$:
$p_{\varphi_k}(t) = {\rm I\!R}e\Big(e^{-i\varphi_k} x(t)\Big)$
\State \hspace{0.5cm} Extract the maxima of $p_{\varphi_k}(t): \{t_j^k,p_j^k\}$
\State \hspace{0.5cm} Interpolate the set $\{ (t_j^k,p_j^k)\}$ to obtain the ''tangent`` in direction $\varphi_k: e_{\varphi_k}^{'}(t)$
\State \textbf{end for}
\State Compute the mean of all tangents: $m(t) = \frac{2}{N}\sum\limits_k e_{\varphi_k}^{'}(t)$ 
\State $d(t) \gets \Tilde{x}(t)-m(t)$ 
\State $\Tilde{x}(t) \gets d(t)$ go to line 3; \textbf{repeat} until $d(t)$ becomes an IMF \Comment{Stoppage Criteria}
\State \textbf{repeat} lines 3:10 until desired number of IMFs obtained 
\EndProcedure
\end{algorithmic}
\end{algorithm}

\subsubsection{Multivariate Empirical Mode Decomposition}
While BEMD addresses the misalignment of IMFs for a given pair of signals, it does not guarantee that the IMFs will be aligned globally if applied in a pair-wise manner to the components of a multivariate signal consisting of more than two signals.  To ensure that the same number of IMFs are produced and to avoid the misalignment of the IMFs when analyzing more than two signals, multivariate extensions of EMD have been proposed.  For example, an extension to the trivariate case was proposed by \cite{ur2010empirical} and a more general multivariate case by \cite{rehman2010multivariate}.  

The latter extension involves processing the signals directly in multidimensional space. First, the signal is projected along different directions in $n$-dimensional space. Using the projections, multiple $n$-dimensional envelopes are computed, which are used to obtain the local mean. Calculation of the local mean can be thought of as an approximation of the integral of all envelopes in multiple directions in $\mathbb{R}^n$.  The accuracy of this integral is bounded by how uniformly the direction vectors used to perform the projections are selected. The direction vectors in $n$-dimensional space can be represented by points on a multi-dimensional sphere of unit radius. 
Hence, the problem becomes finding a uniform sampling scheme on the sphere. This can be obtained either using uniform angular sampling or sampling based on low-discrepancy point sets.  While the first scheme provides a simple and easy way for sampling a unit sphere in $\mathbb{R}^n$, it does not provide a uniform sampling distribution as it gives rise to higher density closer to the poles.  Thus, the second sampling scheme, which relies on the concept of discrepancy (a measure of the non-uniformity of a distribution) to create a uniform point set on spheres is leveraged. A standard approach for obtaining such a point set is based on the Hammersley sequence \cite{cui1997equidistribution}.

Given the set of direction vectors obtained using the aforementioned sampling scheme, the projections $p_{\varphi_k}(t)$, for $t=1 \ldots T$,  of the input signal $\mathbf{x}(t)$ along directions  $\mathbf{v}(t)^{{\varphi_k}}$ in multidimensional space are calculated as $p_{\varphi_k}(t) =\mathbf{x}(t) \mathbf{v}(t)^{{\varphi_k}}$. Thereafter, the time instants $\{t_j^{\varphi_k}\}$ corresponding to the maxima of the projection $p_{\varphi_k}(t)$ are determined.  These extrema are interpolated via cubic spline interpolation to get a series of multivariate signal envelopes $\mathbf{e}^{\varphi_k}(t)$. 
The local mean is computed by averaging the envelopes. This is followed by the sifting process to extract the IMFs from the multivariate data until the stoppage criterion is met for all the projected signals.  The implementation of MEMD \cite{rehman2010multivariate, huang2014hilbert} is shown in Algorithm \ref{alg:memd}.

\begin{algorithm}
\caption{Multivariate Empirical Mode Decomposition}\label{alg:memd}
\begin{algorithmic}[1]
\Procedure{MEMD($\mathbf{x}(t)$, stoppage criteria, number of IMFs)}{}\Comment{$\mathbf{x}(t)$ is the input signal}
\State $\mathbf{\Tilde{x}}(t) \gets\mathbf{x}(t)$
\State Choose the point set based on the Hammersley sequence for sampling on an $(n-1)-$sphere \cite{rehman2010multivariate}
\State \textbf{for} $1\leq k \leq N$ \textbf{do}
\State \hspace{0.5cm} Compute a projection, $p_{\varphi_k}(t)\big{|}_{t=1}^T$ of input signal $\mathbf{x}(t)\big{|}_{t=1}^T$ along the direction vector $\mathbf{v}(t)^{{\varphi_k}}$: $p_{\varphi_k}(t) =\mathbf{x}(t) \mathbf{v}(t)^{{\varphi_k}}$
\State \hspace{0.5cm} Find time instants $\{t_j^{\varphi_k}\}$ corresponding to the maxima of $p_{\varphi_k}(t)$
\State \hspace{0.5cm} Interpolate the set $\{ (t_j^{\varphi_k},\mathbf{x}\big(t_j^{\varphi_k}\big))\}$ to obtain the multivariate envelope  $\mathbf{e}^{\varphi_k}(t)$
\State \textbf{end for}
\State Compute the mean $\mathbf{m}(t)$ of the envelope as: 
$\mathbf{m}(t) = \frac{1}{N}\sum\limits_{k=1}^{N}\mathbf{e}^{\varphi_k}(t)$ 
\State $\mathbf{d}(t) \gets\mathbf{\Tilde{x}}(t)-\mathbf{m}(t)$ 
\State $\mathbf{\Tilde{x}}(t) \gets\mathbf{d}(t)$ go to line 3; \textbf{repeat} until $\mathbf{d}(t)$ satisfies the \Comment{Stoppage Criterion} for a multivariate IMF
\State \textbf{repeat} line 2:11 until desired number of IMFs obtained 
\EndProcedure
\end{algorithmic}
\end{algorithm}

\subsubsection{Noise-assisted Multivariate Empirical Mode Decomposition}

While MEMD is a powerful tool, it can suffer from a mode-mixing phenomena when analyzing real signals.  This refers to when an IMF has components from multiple frequencies \cite{xu2019causes}, which occurs when oscillations with different time scales coexist in the same IMF or when the oscillations in the same time scale are assigned to different IMFs. 
Mode mixing can occur when the frequencies of the constituent signals are too close to one another or when the amplitude of the low frequency signal is too large \cite{xu2019causes}. The issue of mode mixing hampers the application of EMD and MEMD \cite{ur2011filter, gao2008analysis, wu2009ensemble, rilling2007one, xu2019causes}, and to mitigate this issue noise-assisted MEMD (na-MEMD) was introduced \cite{wu2009ensemble,yeh2010complementary, colominas2012noise, zhang2017noise}.  

This approach is based on the fact that when applied to Gaussian WN, both standard and multivariate EMD behave as a filter bank of overlapping bandpass filters where the frequency bands have half the width of their predecessors.  In na-MEND, $p$ uncorrelated Gaussian WN signals of the same length as the original $n$ input signals are created.  By processing the $p+n$ multivariate signals using the MEMD algorithm, multivariate IMFs can be extracted.  Because the added noise signals occupy a broad range in the frequency spectrum,  MEMD aligns the IMFs based on the dyadic filter bank properties described above, which in turn mitigates mode-mixing \cite{ur2011filter, yeh2010complementary}.  The $p$ IMFs corresponding to the added noise signals are discarded, leaving the IMFs corresponding to the original input signal.

\subsection{Multivariate Variational Mode Decomposition}

Variational Mode Decomposition (VMD) is another class of MD techniques. In contrast to EMD, where modes are extracted sequentially, these are non-recursive procedures where the modes are extracted concurrently. They are based on solving an optimization problem where the cost function to be minimized is the sum of bandwidths of all signal modes subject to the constraint that the modes reconstruct the input signal. 

We begin by formulating the univariate VMD problem before extending to the multivariate setting. VMD aims to decompose a real-valued input signal $x(t)$ into $K$ modes, ${u_1,...,u_K}$ where each mode is designed to be compact around a center frequency ${\omega_1,...,\omega_K}$ determined during the course of the decomposition. For each mode, the corresponding analytic signal is computed using the Hilbert transform to obtain the frequency spectrum. The mode's frequency band is shifted to \textit{baseband}, by mixing it with an exponential tuned to the respective center frequency.
The bandwidth is estimated via the squared $L^2$-norm of the gradient.  This constrained problem can thus be formulated as follows:
\begin{equation}
    \min_{\{u_k\},\{\omega_k\}}{ \Bigg\{\sum_k\bigg\| \partial_t\Big[ \Big(\delta(t)+\frac{j}{\pi t} \Big)*u_k(t)\Big]e^{-j\omega_kt}   \bigg\|_2^2\Bigg\}}\;\;\;s.t.\;\; \sum_k{u_k} = x(t) \label{eqn: vmd}
\end{equation}
Here 
$\partial_t$
represents the partial derivative with respect to time, the term $\Big[ \Big(\delta(t)+\frac{j}{\pi t} \Big)*u_k(t)\Big]$ is the analytic representation of the signal corresponding to $u_k(t)$, $*$ represents the convolution operator, and $\delta(t)$ the delta function.

To address the constraint $\displaystyle\sum_k{u_k} = x(t)$, \cite{dragomiretskiy2013variational} proposed the use of a quadratic penalty to enforce reconstruction accuracy of the signal in the presence of noise, and Lagrange multipliers $\lambda$ to enforce the constraint strictly.  The augmented Lagrangian can be formulated as follows:
\begin{multline}
    \mathcal{L}(\{u_k\},\{\omega_k\},\lambda)\coloneqq \alpha \sum_k\bigg\| \partial_t\Big[ \Big(\delta(t)+\frac{j}{\pi t} \Big)*u_k(t)\Big]e^{-j\omega_kt}   \bigg\|_2^2 +\\
    \bigg\|x(t)- \sum_k{u_k}\bigg\|_2^2 + \langle \lambda(t),x(t)- \sum_k{u_k(t)} \rangle \label{eqn: Lagvmd}
\end{multline}
where $\alpha$ is a penalty term specifying the relative importance of the first term.

The saddle point of the augmented Lagrangian in (\ref{eqn: Lagvmd}) provides a solution to the original optimization problem expressed in  (\ref{eqn: vmd}).  This can be found using a series of iterative sub-optimizations denoted ``Alternate Direction Method of Multipliers'' (ADMM) \cite{bertsekas1976multiplier, nocedal2006numerical, eckstein2015understanding}.  

Solutions to these sub-optimization problems \cite{dragomiretskiy2013variational} are given by:
\begin{equation}
        \displaystyle{\hat{u}_{k}^{m+1}(\omega) \xleftarrow{} \frac{\hat{x}(\omega) - \sum\limits_{i < k}\hat{u}_{i}^{m+1}(\omega) - \sum\limits_{i > k}\hat{u}_{i}^m(\omega) +\frac{\hat{\lambda}^m(\omega)}{2}}{1+2\alpha(\omega-\omega_k^m)^2}}
    \end{equation}
and
\begin{equation}
        \omega_k^{m+1}  \xleftarrow{} \frac{{\int \limits_{0}^{\infty}\omega |\hat{u}_{k}^{m+1}(\omega)|^2 d\omega}}{{\int\limits_{0}^{\infty}|\hat{u}_{k}^{m+1}(\omega)|^2 d\omega}}
\end{equation}
where the $\hat{\;\;}$ notation refers to the frequency-domain expression for the corresponding variable, and $m$ the iteration. 

The modes and the center frequencies are updated at each iteration accordingly.  The Lagrangian multiplier is also updated using dual ascent for all $\omega \geq 0$ as follows:
\begin{equation}
    \hat{\lambda}^{m+1}(\omega)\xleftarrow{} \hat{\lambda}^{m}(\omega) + \tau \Big( \hat{x}(\omega) - \sum_k \hat{u}_k^{m+1}(\omega)\Big)
\end{equation}\label{eqn: LagrangianMultip}
Here $\tau$ is the update rate of the Lagrange multiplier and controls the rate of convergence.  The higher the value, the faster the convergence. However, this comes at the potential cost of the algorithm getting stuck in a local optimum.  The iteration continues until the convergence criterion is met.

\begin{algorithm}
\caption{Multivariate Variational Mode Decomposition}\label{alg:mvmd}
\begin{algorithmic}[1]

\State {Initialize: {$\{\hat{u}_{k,n}^1\},\,\{\omega_{k}^1\},\,{\hat{\lambda}_n}^1,\, m\xleftarrow{}0$}}
\State \textbf{repeat}
    \State \hspace{0.5cm} $m\xleftarrow{}m+1$
        \State \hspace{1cm} \textbf{for} $k=1:K$ \textbf{do}
        \State \hspace{1.5cm} \textbf{for} $n=1: N$ \textbf{do} 
    \State \hspace{1cm} Update mode $\hat{u}_{k,n}$: $\displaystyle{\hat{u}_{k,n}^{m+1}(\omega) \xleftarrow{} \frac{\hat{x}_n(\omega) - \sum\limits_{i\neq k}\hat{u}_{i,n}(\omega)+\frac{\hat{\lambda}_n^m(\omega)}{2}}{1+2\alpha(\omega-\omega_k^m)^2}}$
        \State \hspace{1.5cm} \textbf{end for}
    \State \hspace{1.0cm} \textbf{end for}
    \State \hspace{1cm} \textbf{for} $k=1:K$ \textbf{do} 
    \State \hspace{2cm} Update center frequency $\omega_k$: $\omega_k^{m+1}  \xleftarrow{} \frac{\sum\limits_{n}{\int \limits_{0}^{\infty}\omega |\hat{u}_{k,n}^{m+1}(\omega)|^2 d\omega}}{\sum\limits_{n}{\int\limits_{0}^{\infty}|\hat{u}_{k,n}^{m+1}(\omega)|^2 d\omega}}$
    \State \hspace{1cm} \textbf{end for}
    \State \hspace{1cm} \textbf{for} $n=1:N$ \textbf{do} 
        \State \hspace{0.5cm} Update $\lambda_n$ for all $\omega \geq0$: $\hat{\lambda}_n^{m+1} (\omega) = \hat{\lambda}_n^{m} (\omega)  + \tau \Bigg( \hat{x}_n (\omega) - \sum\limits_k{\hat{u}_{k,n}^{m+1}(\omega)}  \Bigg)$
    
    \State \hspace{1cm} \textbf{end for}
    
    \State \hspace{0.5cm} \textbf{until} Convergence $\sum\limits_k \sum\limits_n{\frac{|| \hat{u}_{k,n}^{m+1} - \hat{u}_{k,n}^{m}  ||_2^2}{||\hat{u}_{k,n}^{m}||_2^2}} < \epsilon$

\end{algorithmic}
\end{algorithm}

Now that the core algorithm for VMD is described, the multivariate extension can be more easily understood.  Let $\mathbf{x(t)}$ denote the multivariate input signal $\mathbf{x(t)} = [x_1(t),\,x_2(t),\,..., x_n(t)]$.  The aim is to extract an ensemble of multivariate modes $\{\mathbf{u}_k(t)\}_{k=1}^{K}$ such that the sum of bandwidth of obtained modes is minimum and the sum of the extracted modes can exactly reconstruct the input signal.
Now the corresponding augmented Lagrangian function \cite{ur2019multivariate} can be expressed as follows:
\begin{multline}
    \mathcal{L}(\{u_{k,n}\},\{\omega_k\},\lambda_n)\coloneqq \alpha \sum_k \sum_n \bigg\| \partial_t\Big[ \Big(\delta(t)+\frac{j}{\pi t} \Big)*u_{k,n}(t)\Big]e^{-j\omega_kt}   \bigg\|_2^2 +\\
    \sum_n\bigg\|x_n(t)- \sum_k{u_{k,n}}\bigg\|_2^2 + \sum_n\langle \lambda_n(t),x_n(t)- \sum_k{u_{k,n}(t)} \rangle 
\end{multline}\label{eqn: multiLagvmd}

The algorithm for the multivariate extension is shown in Algorithm \ref{alg:mvmd} with steps similar to the univariate VMD described above. Both univariate and multivarite VMD are shown to be robust to noise and the influence of sampling, and able to effectively separate frequency components. MVMD also enjoys the mode alignment properties for multivariate signals, and additionally shows superior filter bank properties compared to MEMD and is able to effectively extract quasi-orthogonal modes \cite{dragomiretskiy2013variational, ur2019multivariate}. 
Hence, the approach is ideal for our application due to the low signal-to-noise ratio (SNR) in rs-fMRI signal \cite{biswal1996reduction, tanabe2002comparison, garg2013gaussian, drew2020ultra,  teeuw2021reliability}.

\subsubsection{Computing time-varying PS}

In this section, we describe the general framework in which the various MD techniques are used in conjunction with the Hilbert transform to measure time-varying PS in fMRI data.  

The selected MD technique is used to decompose the signal into its IMFs. The IMF of interest is then treated as the input to the Hilbert transform as they are band-limited signal in a manner that respects Bedrosians theorem. This allows the phase of the IMFs to be estimated and used in PS analysis.

When performing time-varying PS analysis, there exist two classes of techniques WPS and IPS.  WPS methods use metrics that provide a single omnibus measure of the PS between two signals. A sliding window technique is used to compute the metric locally within a specific time window. As the window is shifted across time, one can obtain a time-varying value of the measure of interest (i.e., the synchronization between the two signals). The types of metrics that can be used to measure WPS are discussed in \cite{honari2021evaluating}. In this work, we use circular-circular correlation which is an extension of the standard correlation coefficient for angular values.

In the WPS framework, it is common to use a boxcar as a window \cite{honari2021evaluating, rebollo2018stomach}. However, similar to its use in the CSW framework, it has been shown that tapering leads to smoother results \cite{shakil2017parametric} and alleviates the effect of sudden changes related to the edges of rectangular windows.  Further, when using a boxcar window all points in the window are given equal weight, which in turn leads to inflated sensitivity to outliers \cite{allen2014tracking, mokhtari2019sliding}.  To alleviate these concerns, we propose the use of a tapered window based on the von-Mises probability density function (pdf) which we refer to as tapered-WPS (tWPS) \cite{honari2021measuring}.  

The pdf for a random circular variable in the range $(-\pi,\pi]$ that follows the von-Mises distribution with circular mean $\Theta$ and variance $1/\kappa$ is given by:
\begin{equation}
    f(\theta | \Theta, \kappa) = \frac{1}{2\pi I_0(\kappa)} e^{\kappa cos(\theta - \Theta)} 
    \label{Eq: vMpdf}
\end{equation}
where $\kappa \geq 0$ and $I_0$ is the zero-order modified Bessel function of the first kind. 

\begin{figure}[htp]
\begin{center}
\includegraphics[width=\textwidth,angle =0,trim={0cm 0cm 0cm 0cm},clip]{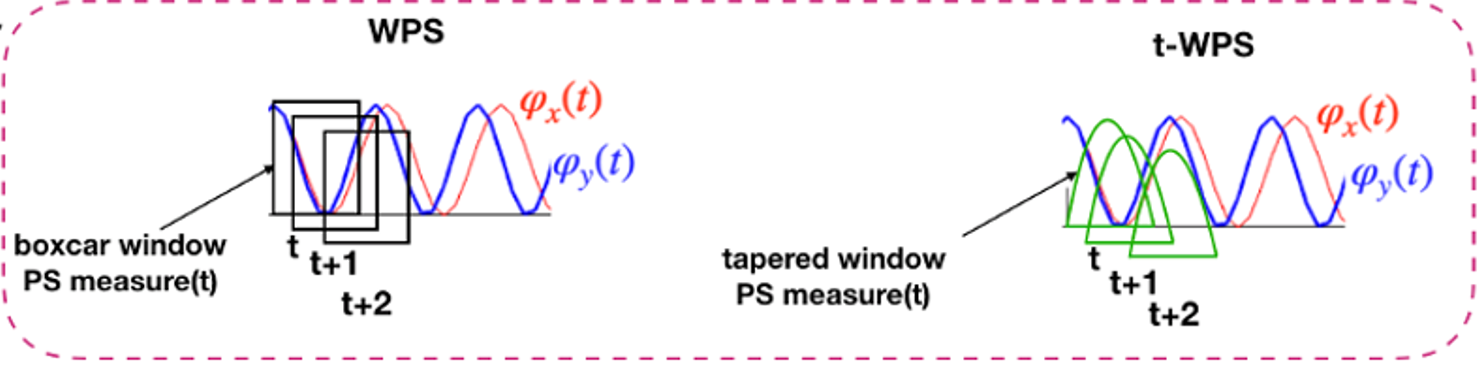}
\end{center}
\caption{A comparison of WPS and tWPS for the analysis of time-varying PS. (Left) WPS using a rectangular boxcar window and (right) tWPS using a von-Mises window.}\label{fig:tWPS}
\end{figure}

Based on the chosen window size, the pdf is discretized into an equivalent number of equidistant samples to create the window.  A schematic comparison of WPS and tWPS is shown in Figure \ref{fig:tWPS}.

In contrast, IPS methods directly analyze the instantaneous phases extracted using the Hilbert Transform.  The benefit of using an IPS approach is that it offers single time-point resolution of time-resolved fMRI connectivity, and does not require choosing an arbitrary window length as is the case for WPS methods.  \cite{honari2021evaluating} proposed the cosine of the relative phase (CRP) as a measure of IPS. 
It is defined as follows:
\begin{equation}
    \vartheta(t) = \cos{(\Delta\Phi(t))}
\end{equation}
where $\Delta\Phi(t))$ represents the phase difference between the signals at time $t$. CRP was shown to outperform other IPS measures and overcome issues related to undetected temporal transitions from positive to negative associations common in IPS analysis.  Further, CRP unfolds the distribution of PS measures as opposed to another popular approach Phase Coherence. This is beneficial when performing  subsequent clustering of PS matrices in the analysis.  In this paper, we use CRP to measure time-varying PS.

\subsection{Simulations}\label{sec: simulation}

To evaluate the performance of the different MD techniques in the context of PS analysis we perform a series of simulations.
For each simulation we combined the different MD techniques with the Hilbert transform to compute the phase. The three MD techniques used throughout are BEMD, na-MEMD, and MVMD. Next, time-varying PS measures were computed using tWPS evaluated using circular-circular correlation and a von-Mises window, and IPS evaluated using CRP \cite{honari2021evaluating}.

In the first three simulations, the simulated signal is bivariate.
The first simulation investigates the performance of the three MD techniques in a null setting, while the second and third investigate their performance when two sinusoidal signals have the same frequency, but differing types of phase shifts. 

For all three simulations, data was generated with a sampling frequency of $1/TR$, where $TR$ corresponds to the repetition time of an fMRI experiment. Here we use $TR=2$ seconds.  

In the fourth simulation, a more complex situation is considered. Here a bivariate signal is generated where one of the signals is monocomponent with a central frequency of $f$ and the other signal is multicomponent formed using two frequencies $f$ and $1.1f$.  The simulation allows for the evaluation of each MD techniques ability to isolate the frequency of interest for the use in PS analysis.  

In the fifth simulation, the simulated signal is trivariate. This allows for comparison of applying BEMD in a pair-wise manner with the use of na-MEMD and MVMD which jointly decompose the three signals. After applying the various MD techniques, and using the IMFs corresponding to the central frequency of the simulated signals, an array of time-varying PS measures between all pairs of signals measured at each time point is constructed.  Following the approach of \cite{allen2014tracking}, $k-$means clustering is applied to these arrays to estimate the underlying `brain states' used to generate the signal.

Finally, the sixth simulation aims to compare the performance of the various MD techniques at different noise levels. This simulation mimics the settings from Simulation 3, but alters the amount of noise to investigate its impact on the ability to recover IMFs and use them to evaluate time-varying PS.

All simulations were repeated $1000$ times, and the mean and variance of the time-varying PS measured at each time point was used to construct a $95\%$ confidence interval.  Furthermore, the effect of different window lengths in the WPS analysis was evaluated using three different values ($30$, $60$, and $120$ time points). Below we describe in detail how data was generated for each simulation.

\underline{\em Simulation 1:}  \,\, To simulate signals with independent phase dynamics, we generated two independent random signals from a Gaussian distribution with mean $0$ and standard deviation $1$.  Using the logic of surrogate data testing, we generated surrogate data under the assumption of no relationship between the phase from the two signals. To achieve this goal we used cyclic phase permutation (CPP) surrogates \cite{lancaster2018surrogate}, constructed by reorganizing the cycles within the extracted phase of the signals. This destroys any phase dependence between the pair, while preserving the general form of the phase dynamics of each signal. For this simulation, the $1000$ realizations of signal pairs were generated using CPP surrogates.

\underline{\em Simulation 2:} \,\, Two sinusoidal signals were generated with the same frequency, but with a time-varying phase shift corresponding to a ramp function. To elaborate, consider two sinusoidal signals $x(t)$ and $y(t)$. Let $x(t)$ be the reference signal with an angular frequency of $\omega_0$ and phase $\varphi_x(t)$. Further, let $y(t)$ have the same angular frequency but with phase $\varphi_y(t)$. The signals can be expressed as follows:
\begin{equation}\label{eqn: signal phase shift}
    \begin{split}
        x(t) &= A_x cos\big(\omega_0 t + \varphi_x(t)\big) + \varepsilon_{x}\\
        y(t) &= A_y cos\big(\omega_0 t + \varphi_y(t) \big) + \varepsilon_{y}
    \end{split}
\end{equation}

Without loss of generality, let $\varphi_x(t)=0$ and $\varphi_y(t)$ be a ramp function, 
\begin{equation}\label{eq: rampfunction}
    r(t-t_0) = 
    \begin{cases}
        \begin{aligned}
            &0 \qquad  &t &\leqslant t_0\\
            &t-t_0 \qquad &t &> t_0 
            \end{aligned}
        \end{cases}
\end{equation}
Throughout, we set $\omega_0 = 2\pi f \;rad/s$ with $f=0.05 \, Hz$, $A_x=A_y=1$ and set the transition to occur at $t_0 = 170 \, s$.  The noise terms $\varepsilon_{x}$ and $\varepsilon_{y}$ are independent Gaussian white noise with mean $0$ and standard deviation $1$.  
Under this formulation, the two signals start out phase synchronized and remain in this state up to $t_0 = 170 \, s$.  After this time the phase difference starts linearly increasing and transitioning into a non-synchronized state.

{\em Simulation 3:} \, \,  Two sinusoidal signals were generated with the same frequency, but with a time-varying phase shift corresponding to a sigmoid function. As in the previous simulation, data was generated according to (\ref{eqn: signal phase shift}). Here  $\varphi_x(t)=0$ and $\varphi_y(t)$ is a sigmoid function, i.e. 
\begin{equation}
    s(t-t_0) = \frac{a}{1+\exp{\big(b(t-t_0)\big)}}
\end{equation}
Throughout, we set $a=2\pi$, $b=-0.01$, $t_0 =170$, and $\omega_0 = 2\pi f \;rad/s$ with $f=0.05 \, Hz$ and $A_x=A_y=1$. The noise terms $\varepsilon_{x}$ and $\varepsilon_{y}$ are independent Gaussian white noise with mean $0$ and standard deviation $1$.  
Under this formulation, the signals are initially in phase, after which the amount of phase shift gradually increases. This continues until $t=170$ when the pairs are in anti-phase synchronization. Thereafter, the signals gradually return to being in phase.  The transition between the phase of the signals from 0 to $2\pi$ occurs smoothly and monotonically increasing. 

{\em Simulation 4:} \, \, A bivariate signal was generated where the phase shift between the signals $x(t)$ and $y(t)$ vary as a sigmoid function as in Simulation 3.  However, here the signal $y(t)$ is multicomponent, consisting of two frequencies of $f$ and $1.1f$. They are placed close to one another to assess the performance of the MD techniques in successfully untangling the frequency components from one another and avoid mode mixing.

Hence, the signals can be expressed as follows:
\begin{equation}
    \begin{split}
        x(t) &= A_{x} cos(\omega_0 t) + \varepsilon_{x}\\
        y(t) &= A_{y_1} cos\Bigg(\omega_0 t + \frac{a}{1+\exp{\big(b(t-t_0)\big)}} \Bigg)\\ 
        &\;\;+ A_{y_2} cos\Bigg(1.1\omega_0 t + \frac{a}{1+\exp{\big(b(t-t_0)\big)}} \Bigg)+ \varepsilon_{y}
    \end{split}
\end{equation}
Throughout, we set $a=2\pi$, $b=-0.01$, $t_0 =170$, and $\omega_0 = 2\pi f \;rad/s$ with $f=0.05 \, Hz$ and $A_x=A_{y_1}=A_{y_2}=1$. The noise terms $\varepsilon_{x}$ and $\varepsilon_{y}$ are independent Gaussian white noise with mean $0$ and standard deviation $1$.  

{\em Simulation 5:} \, \, A trivariate signal was generated where the phase shift of each signal varied as follows: 
\begin{equation}
    \begin{split}
        \varphi_{x_1}(t) &= \pi \Big( \Pi_{50,125}(t)  +\Pi_{150,250}(t) + \Pi_{300,400}(t) + \Pi_{500,400}(t) \Big)\\
        \varphi_{x_2}(t) &= \pi \Big( \Pi_{50,125}(t)  -\Pi_{300,400}(t) \Big)\\
        \varphi_{x_3}(t) &= -\pi \Big( \Pi_{150,250}(t)  +\Pi_{300,400}(t) \Big)
    \end{split}
\end{equation}\label{eqn: phase relation}
Here $\Pi_{a,b}(t) = H(t-a) - H(t-b)$ is a boxcar function and $H(t)$ denotes the Heaviside step function (i.e., unit step function). Thus the signals can be expressed as follows:
\begin{equation}
    \begin{split}
        x_1(t) &= A_{x_1} cos\Big(\omega_0 t + \varphi_{x_1}(t)\Big) + \varepsilon_{x_1}\\
        x_2(t) &= A_{x_2} cos\Big(\omega_0 t + \varphi_{x_2}(t)\Big) + \varepsilon_{x_2}\\
        x_3(t) &= A_{x_3} cos\Big(\omega_0 t +\varphi_{x_3}(t) \Big) + \varepsilon_{x_3}
    \end{split}
\end{equation}
Throughout, we set $\omega_0 = 2\pi f \;rad/s$ with $f=0.05 \, Hz$ and $A_{x_1}=A_{x_2}=A_{x_3}=1$. The noise terms $\varepsilon_{x_1}$, $\varepsilon_{x_2}$ and $\varepsilon_{x_3}$ are independent Gaussian white noise with mean $0$ and standard deviation $1$. 

\begin{figure}[htp]
\begin{center}
\includegraphics[width=0.9\textwidth,trim={0cm 7cm 0cm 2cm},clip]{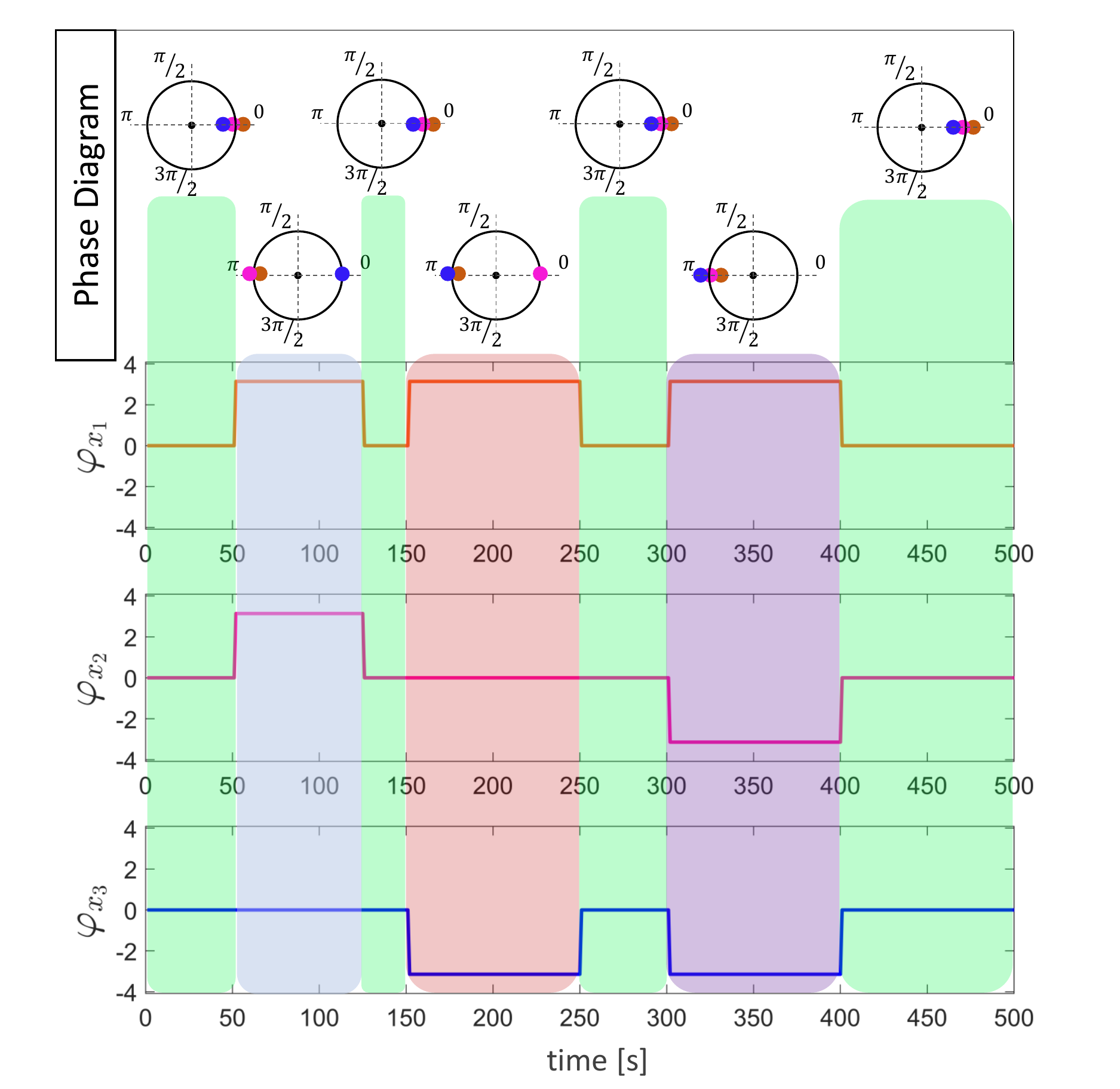}
\end{center}
\caption{The ground truth phase used to generate the trivariate signals $(x_1, \, x_2, \, x_3)$ in Simulation 5. The phase of each signal are also shown in a phase diagram for each epoch on the unit circle. Note the green and purple intervals both illustrate a situation where all signals are in-phase (State 2).  }\label{fig: Sim5_grndtruthphase}
\end{figure}

The relationship between the signals is illustrated in Figure \ref{fig: Sim5_grndtruthphase}.  The three signals are initially in phase up to $50\,[s]$, after which $x_1(t)$ and $x_2(t)$ transition to being in anti-phase with $x_3(t)$, while remaining in phase with one another. This relationship continues up to $t = 125\,[s]$, after which all three signals are again in phase. At $t = 150\,[s]$ the phase of $x_2$ is in anti-phase with $x_1(t)$ and $x_3(t)$, while $x_1(t)$ and $x_3(t)$ are in phase (i.e., $\varphi_{x_1}-\varphi_{x_3} = \pi - (\-pi) = 2\pi$).  From $t= 250\,[s]$ to $t= 300\,[s]$ all three signals are again in phase.  From $t= 300\,[s]$ to $t= 400\,[s]$ $\varphi_{x_1} = \pi$ and $\varphi_{x_2} = \varphi_{x_3} = - \pi$ and thus all three signals are in phase.  Finally, at $t=400$ the phase of all three signals transition to 0.  

In total there are three states that the signals transition between. In the time period $[150, \, 250]$ the signals are in State 1. Here $x_1(t)$ and $x_3(t)$ are in anti-phase with  $x_2$, while in phase with one another.   In the time periods $[0, \, 50]$, $[125, \, 150]$, and $[250, \, 500]$ the signals are in State 2. Here all signals are in-phase. In the time period $[50, \, 125]$ the signals are in State 3. Here $x_1(t)$ and $x_2(t)$ are in anti-phase with $x_3(t)$, while in phase with one another. For each repetition of the simulation, we seek to determine whether we can recreate these  states using a variant of the approach proposed by \cite{allen2014tracking}.  First, for each combination of MD technique and PS measure, an array of time-varying PS between all pairs of signals measured at each time point are constructed.  Next, $k-$means clustering is applied to these arrays to estimate the underlying `brain states' used to generate the signal. 

{\em Simulation 6:} \, \, A bivariate signal was generated where the phase shift between the signals $x(t)$ and $y(t)$ vary as a sigmoid function as in Simulation 3.  However, in this simulation the performance of each of the MD techniques are studied under various noise levels. Here we let $Var(\varepsilon_{x}) = Var(\varepsilon_{y}) = 1,\,4,\,10$, where the noise are independent Gaussian noise with mean 0.

\subsection{Analysis of HCP Data}

In this section, the methods described above are applied to rs-fMRI data from the Human Connectome Project. The data is  briefly described below.

\subsubsection{Image Acquisition}
The 2014 Human Connectome Project 500 Parcellation+Timeseries+Netmats (HCP500- PTN) release \cite{van2013wu} is a collection of neuroimaging data from $523$ healthy adults acquired on a customized 3T Siemens connectome-Skyra 3T scanner. Participants completed two scanning sessions on two separate days. A T1w MPRAGE structural run was acquired during each session (acquisition time $= 7.6\; min$, TR/TE/TI = $2400/2.14/1000\;ms$, resolution = $0.7 \times 0.7 \times 0.7\; mm^3$, SENSE factor $= 2$, flip angle $= 8 \degree$). A simultaneous multi-slice pulse sequence with an acceleration factor of eight \cite{uugurbil2013pushing} was used to acquire two rs-fMRI runs during each session, each consisting of $1200$ volumes sampled with TR $0.72\;s$, at $2\;mm$ isotropic spatial resolution (TE $= 33.1\;ms$, flip angle $= 52 \degree$, $72$ axial slices). Participants were instructed to keep their eyes open and fixated on a cross hair, while remaining as still as possible. Within sessions, phase encoding directions for the two runs were alternated between right-to-left (RL) and left-to-right (LR) directions.  In addition, respiratory signals associated with each scan where acquired using a respiratory belt placed on the participant's abdomen.

\subsubsection{Preprocessing}
We used the preprocessed and artifact-removed rs-fMRI data provided through the HCP500-PTN data release. The preprocessing and the artifact-removal procedures used are described elsewhere \cite{glasser2013minimal,smith2013resting, griffanti2014ica, salimi2014automatic}. Briefly, each run was minimally preprocessed \cite{glasser2013minimal,smith2013resting}, and artifacts removed using the ICA-based X-noiseifier (ICA + FIX) procedure \cite{griffanti2014ica,salimi2014automatic}. The rs-fMRI data from each run were represented as a time series of grayordinates, a combination of cortical surface vertices and subcortical standard-space voxels \cite{glasser2013minimal}. Each run was temporally demeaned and variance normalized \cite{beckmann2004probabilistic}. All four runs for 461 participants were fed into MELODIC's Incremental Group-Principal Component Analysis (MIGP) algorithm, which estimated the top 4500 weighted spatial eigenvectors. GICA was applied to the output of MIGP using FSL's MELODIC tool \cite{beckmann2004probabilistic} using five different dimensions (25, 50, 100, 200, 300). In this study, we used data with dimension d = 100 to perform further analysis. Dual-regression was used to map group-level spatial maps of the components onto each subject's time series data \cite{filippini2009distinct}. The full set of group-level maps were used as spatial regressors against each subject's full time series (4800 volumes) to obtain a single representative time series per independent component.

 The respiratory node was sampled at $400\;Hz$ while the 100 brain nodes were sampled at $1/TR$ with $TR=0.72 s$.  The respiratory node was therefore resampled by applying an anti-aliasing low-pass filter to the signal and compensating for the delay introduced by the filter.  This is done by zero padding the signal, applying a finite impulse response  anti-aliasing filter to up-sample the input signal, and discarding samples to down-sample the filtered signal by ($400/(1/0.72)$) = 288.  This was performed in MATLAB using the 'resample' function, which uses a Kaiser window.  

\subsubsection{Analysis}
Data from a single run of the HCP data, consisting of 100 ROIs and 1200 time points per subject, was used.  The downsampled respiration signal was also included to investigate its relation with the rs-fMRI signal.  In total the analysis was performed on 101 nodes. 

We applied MVMD to data from 50 participants,  decomposing the multivariate time series into 10 IMFs.  We choose two IMFs for further analysis. The first corresponds to the highest power spectrum in the range $0.01-0.1\,Hz$, which is the standard  frequency band used in rs-fMRI studies. The second corresponds to the highest power spectrum close to the respiration frequency. The latter will allow for the evaluation of PS between brain and respiration.

Applying the Hilbert transform to the modes allowed us to compute the instantaneous phase for the 101 nodes across time.
For each mode and each pair of participant-specific signals, we 
computed time-varying PS using CRP.
This gave rise to a $101 \times 101 \times 1200$ array of IPS measures.  Thus, at each time point a $101 \times 101$ matrix analogous to the correlation matrix used in the sliding window technique was constructed.   

Brain states were estimated using the approach proposed by \cite{allen2014tracking}.  This was done by first reorganizing the lower triangular portion of each participant's data into a matrix with dimensions $5050 \times 1200$; here the row dimension corresponds to the number of elements in the lower triangular portion of the matrix (i.e., $101(101-1)/2$), and the column dimension to the number of time points. The data from all participants was concatenated into a matrix with row dimensions $5050$ and column dimensions $(1200 \times 50 = 60000)$. Finally, $k$-means clustering was applied to the concatenated data set, where each of the resulting cluster centroids were assumed to represent a recurring brain state.  In this study, we chose the number of centroids to equal two, representing two distinct brain states.  This value was chosen based on the Davies-Bouldin Index (DBI)\cite{davies1979cluster}, which is a measure of clustering quality.

\section{Results}
\subsection{Simulations}

\begin{figure}[tp]
\begin{center}
\includegraphics[width=\textwidth,trim={0cm 7.5cm 0cm 7cm},clip]{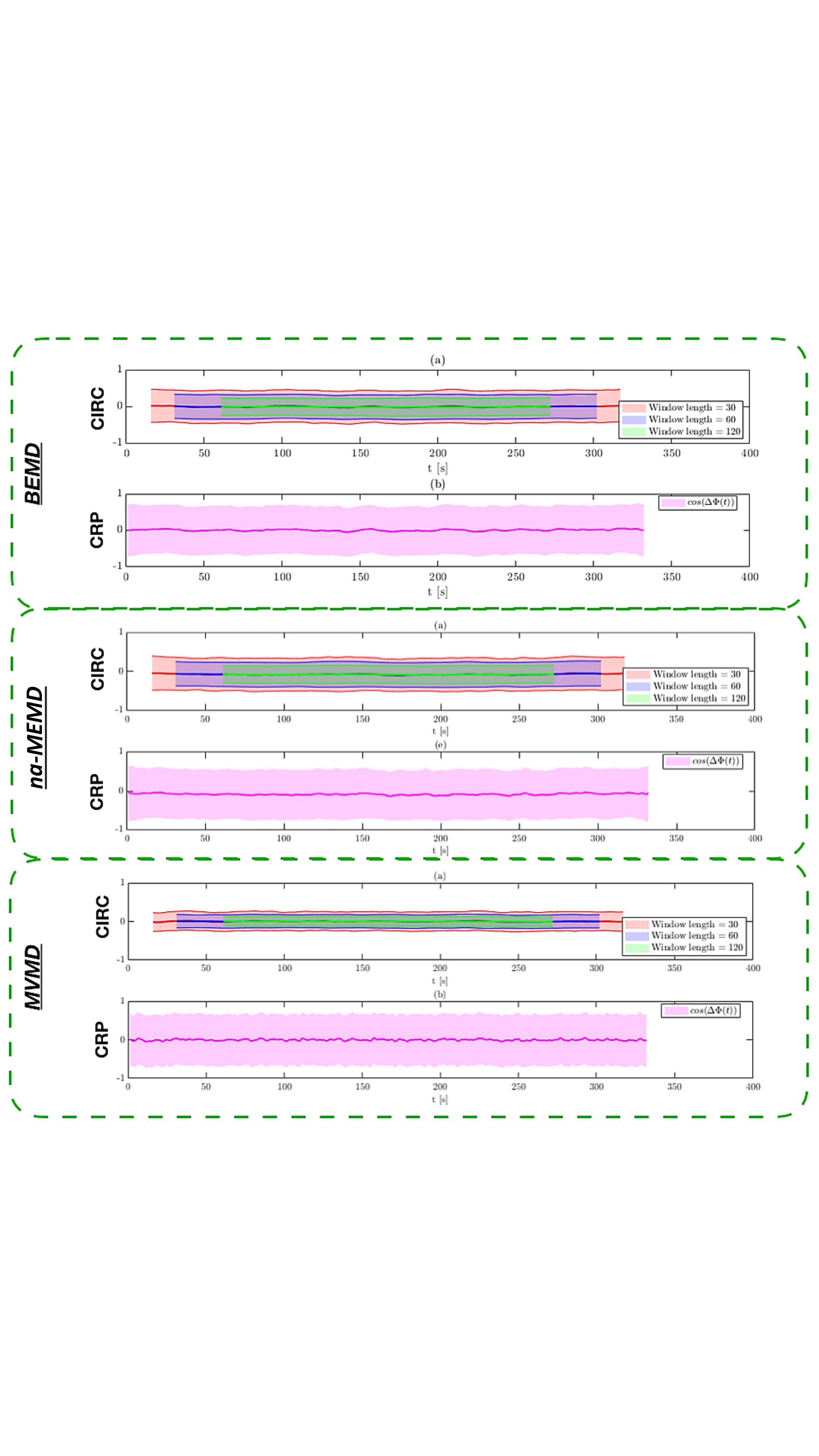}
\end{center}

\caption{Results of Simulation 1. A comparison of BEMD, na-MEMD, and MVMD-based PS under the the null setting. Results are based on an IMF extracted with a central frequency of $0.05 \, Hz$. Results are shown for tapered WPS using circular-circular correlation (CIRC; panels (a), (c), (e)) and IPS using cosine of relative phase (CRP; panels (b), (d), (f)).  In each panel, the mean and 95\% interval for each measure are shown at each time point. For the tWPS measures results are shown for three different window lengths (30, 60, and 120).  }
\label{fig: Sim1_PSMeasures1000}
\end{figure}

\begin{figure}[!htp]
\begin{center}
\includegraphics[width=\textwidth,trim={0cm 4cm 0cm 9cm},clip]{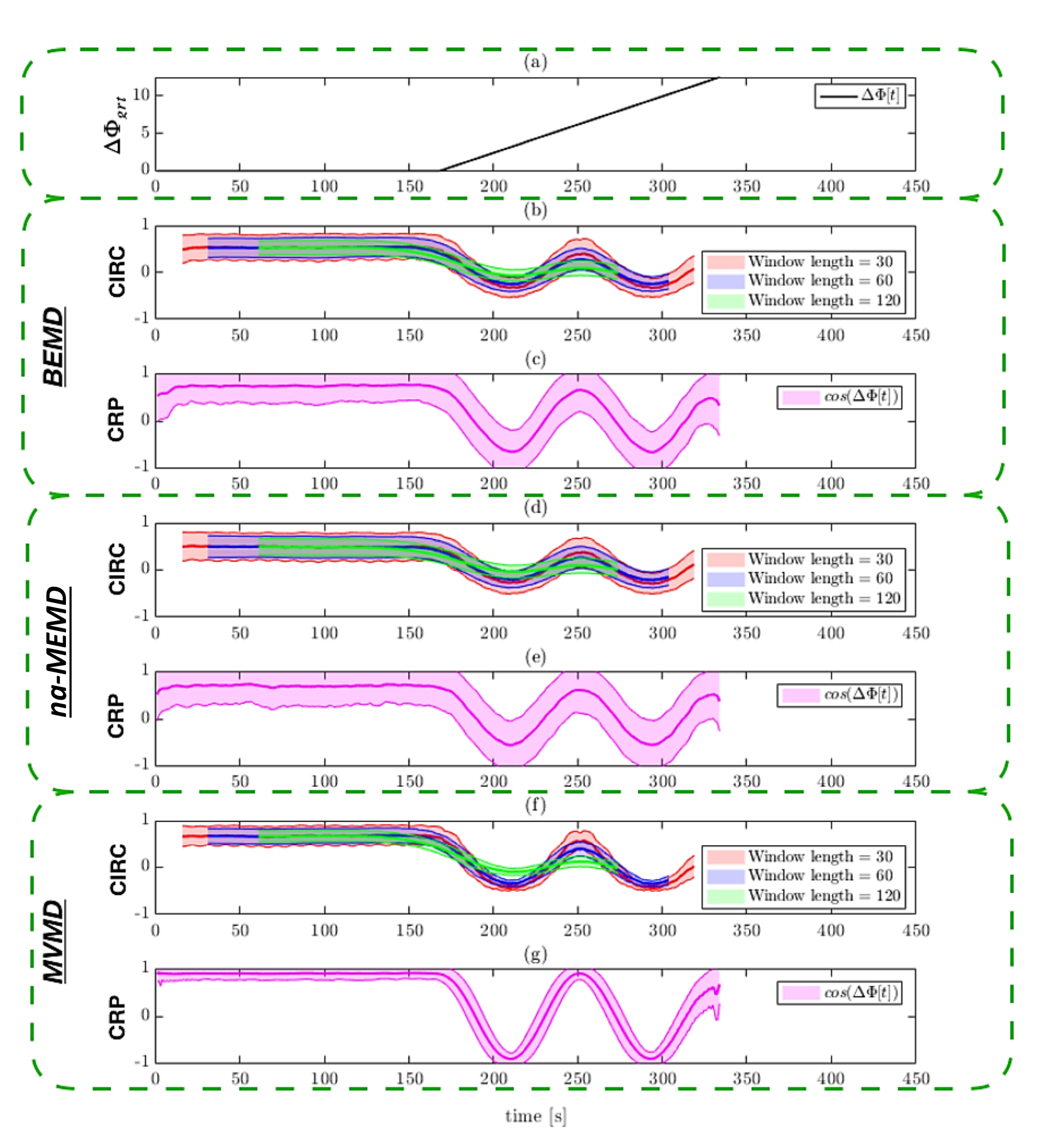}
\end{center}
\caption{ Results of Simulation 2. A comparison of BEMD, na-MEMD, and MVMD-based PS when the phase shift corresponds to a ramp function (panel (a)). Results are based on an IMF extracted with a central frequency of $0.05 \, Hz$. Results are shown for tWPS using circular-circular correlation (CIRC; panels (b), (d), (f)) and IPS using cosine of relative phase (CRP; panels (c), (e), (g)).  In each panel, the mean and 95\% interval for each measure are shown at each time point. For the tWPS measures results are shown for three different window lengths (30, 60, and 120).   }\label{fig: Sim2_PSMeasures1000}
\end{figure}

\begin{figure}[htp]
\begin{center}
\includegraphics[width=\textwidth,trim={0cm 3cm 0cm 11.1cm},clip]{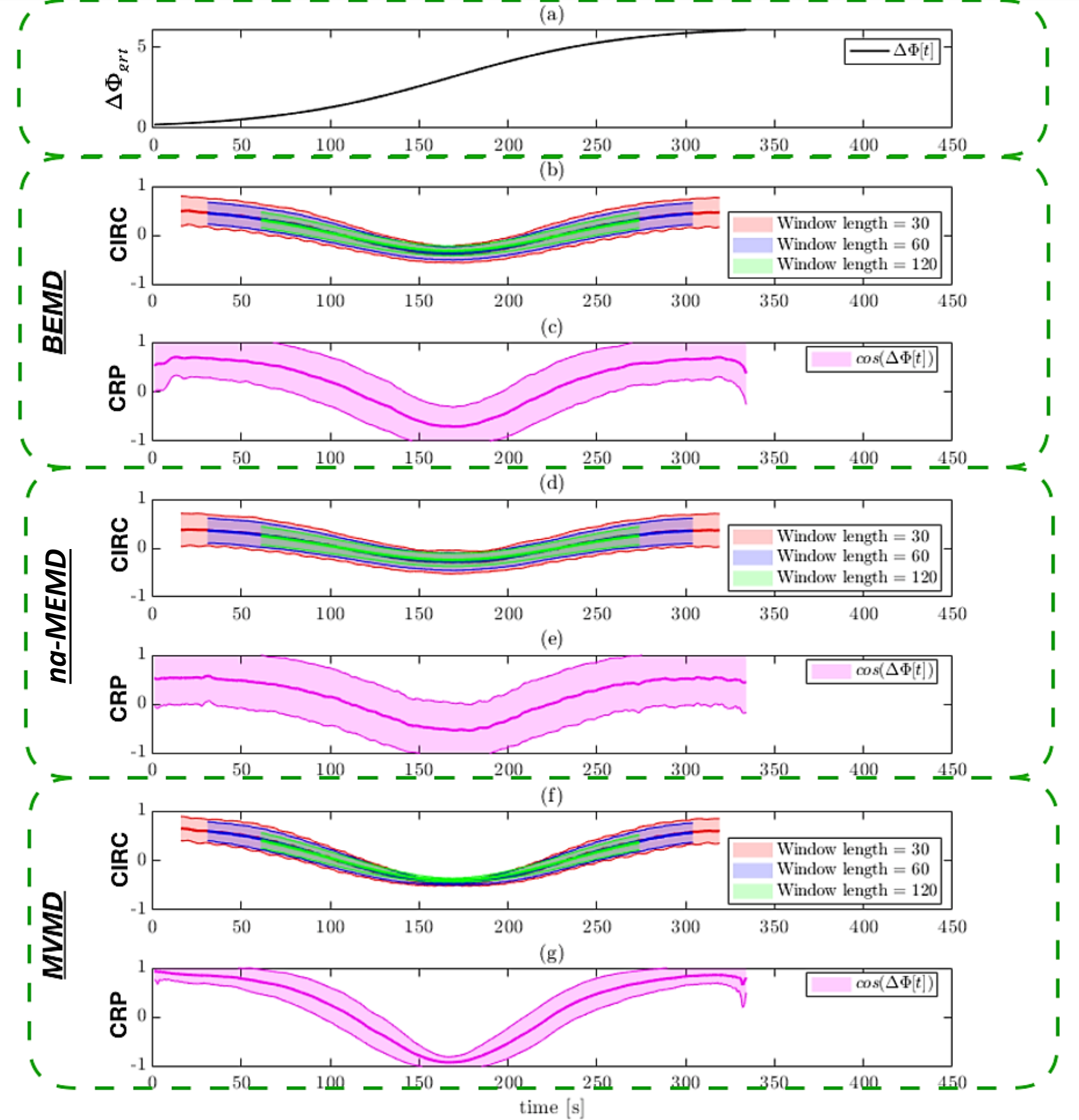}
\end{center}
\caption{Results of Simulation 3. A comparison of BEMD, na-MEMD, and MVMD-based PS when the phase shift corresponds to a sigmoid function (panel (a)). Results are based on an IMF extracted with a central frequency of $0.05 \, Hz$. Results are shown for tWPS using circular-circular correlation (CIRC; panels (b), (d), (f)) and IPS using cosine of relative phase (CRP; panels (c), (e), (g)).  In each panel, the mean and 95\% interval for each measure are shown at each time point. For the tWPS measures results are shown for three different window lengths (30, 60, and 120).  }\label{fig: BEMD_Sim3_PSMeasures1000}
\end{figure}

\begin{figure}[!htp]
\begin{center}
\includegraphics[width=1\textwidth,trim={0cm 4.5cm 0cm 1cm},clip]{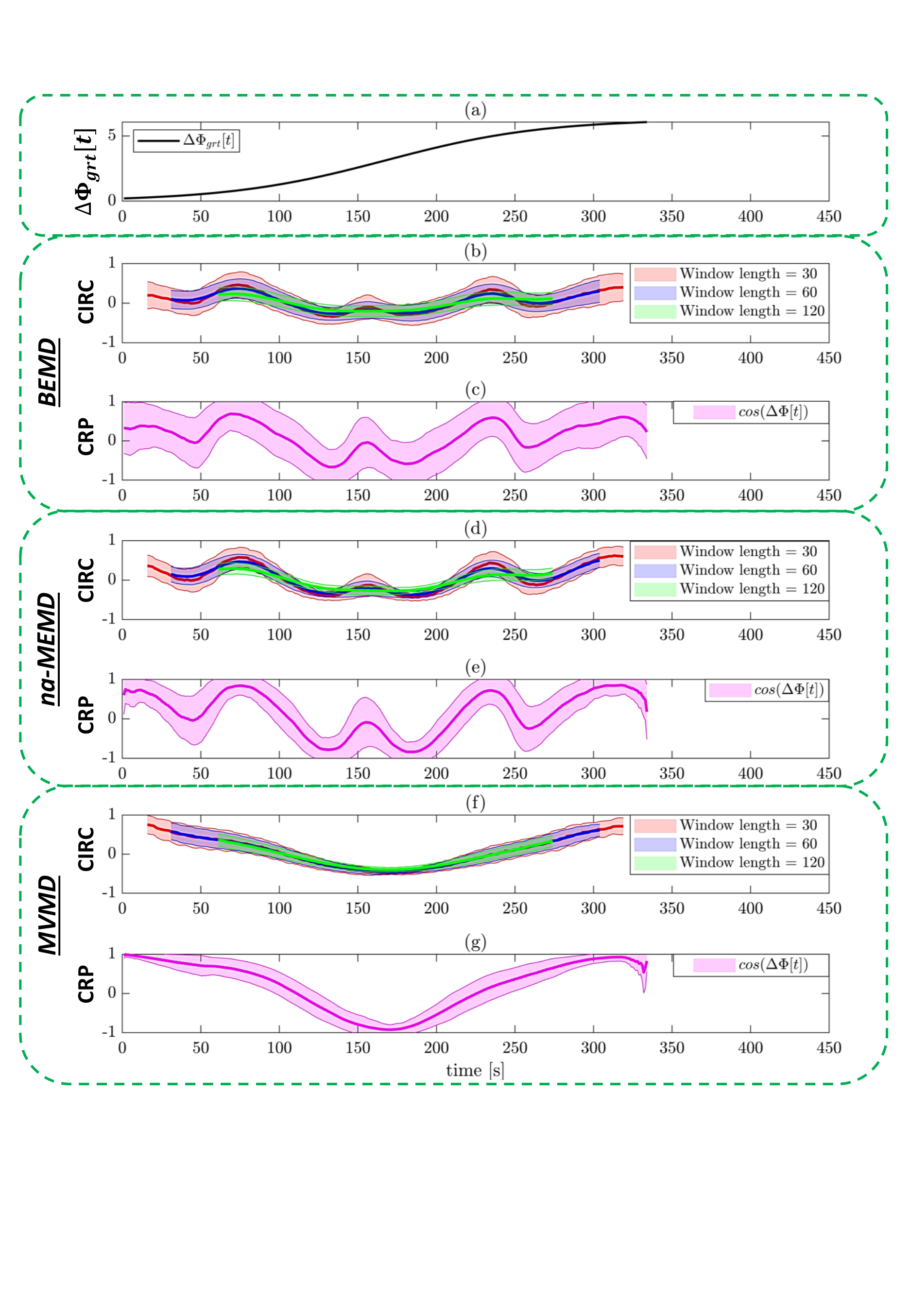}
\end{center}
\caption{Results of Simulation 4. A comparison of BEMD, na-MEMD, and MVMD-based PS when one of the signals is multicomponent.  The simulation compares the performance of MD approaches to handle mode mixing.  The true phase shift corresponds to a sigmoid function (panel (a)). Results are based on an IMF extracted with a central frequency of $0.05 \, Hz$. Results are shown for tWPS using circular-circular correlation (CIRC; panels (b), (d), (f)) and IPS using cosine of relative phase (CRP; panels (c), (e), (g)).  In each panel, the mean and 95\% interval for each measure are shown at each time point. For the tWPS measures results are shown for three different window lengths (30, 60, and 120).  }\label{fig: Sim4_PSMeasures}
\end{figure}

\begin{figure}[htp]
\begin{center}
\includegraphics[width=0.97\textwidth,trim={0cm 1.5cm 0cm 0.9cm},clip]{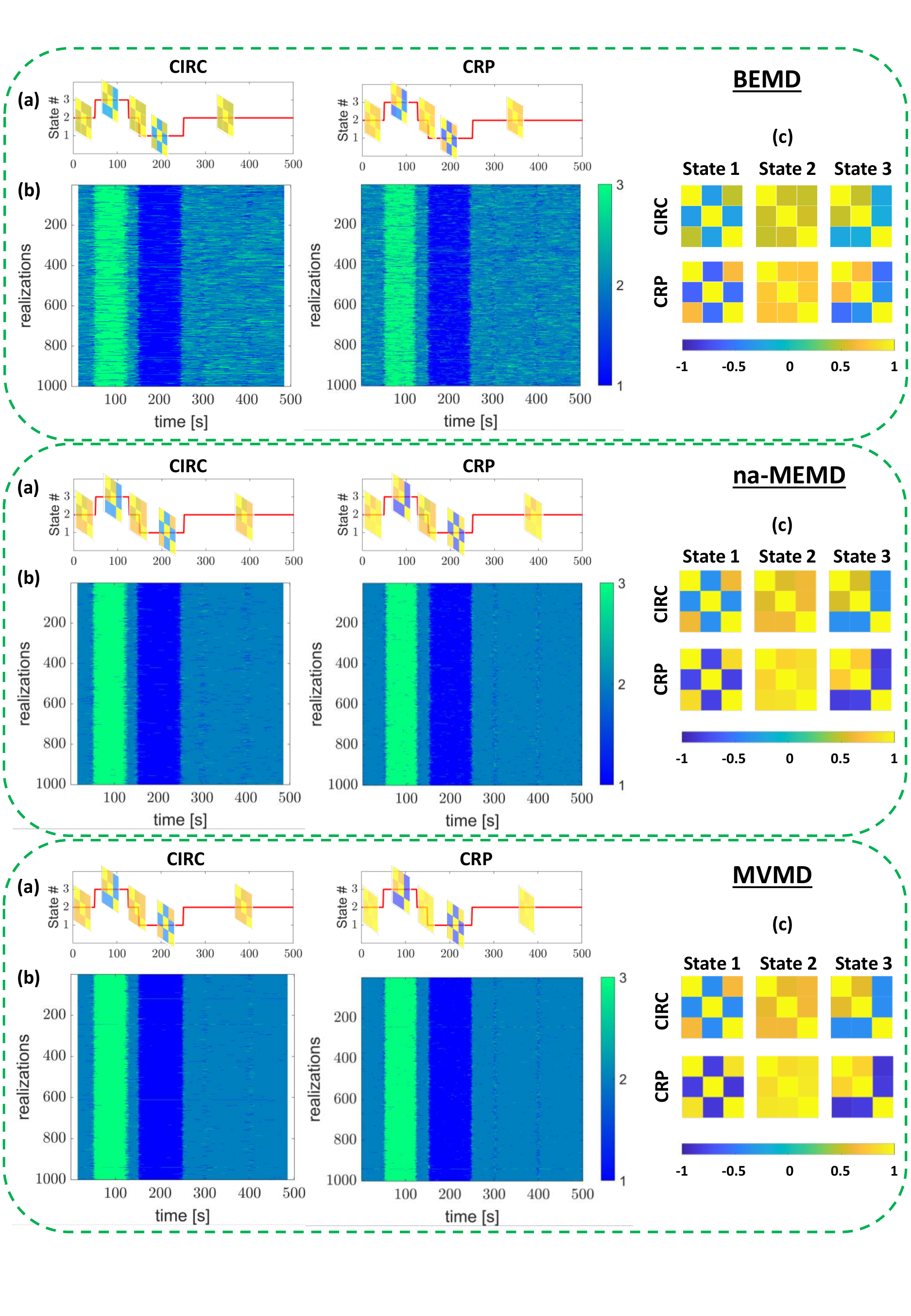}
\end{center}
\caption{Results of Simulation 5. Comparison of BEMD, na-MEMD, and MVMD-based PS applied to a multivariate signal.  For each of the three MD approaches, panel (a) shows the ground truth state transition along the estimated state overlaid on each state; panel (b) displays a heat map of the estimated state transitions across time for the 1000 realization for each PS measure; and panel (c) shows the average estimated states across the 1000 realizations for each PS measure. }\label{fig: multivariate_Sim4_PSMeasures}
\end{figure}

\begin{figure}[!ht]
\begin{center}
\includegraphics[width=\textwidth,trim={0.3cm 10.5cm .2cm 8.5cm},clip]{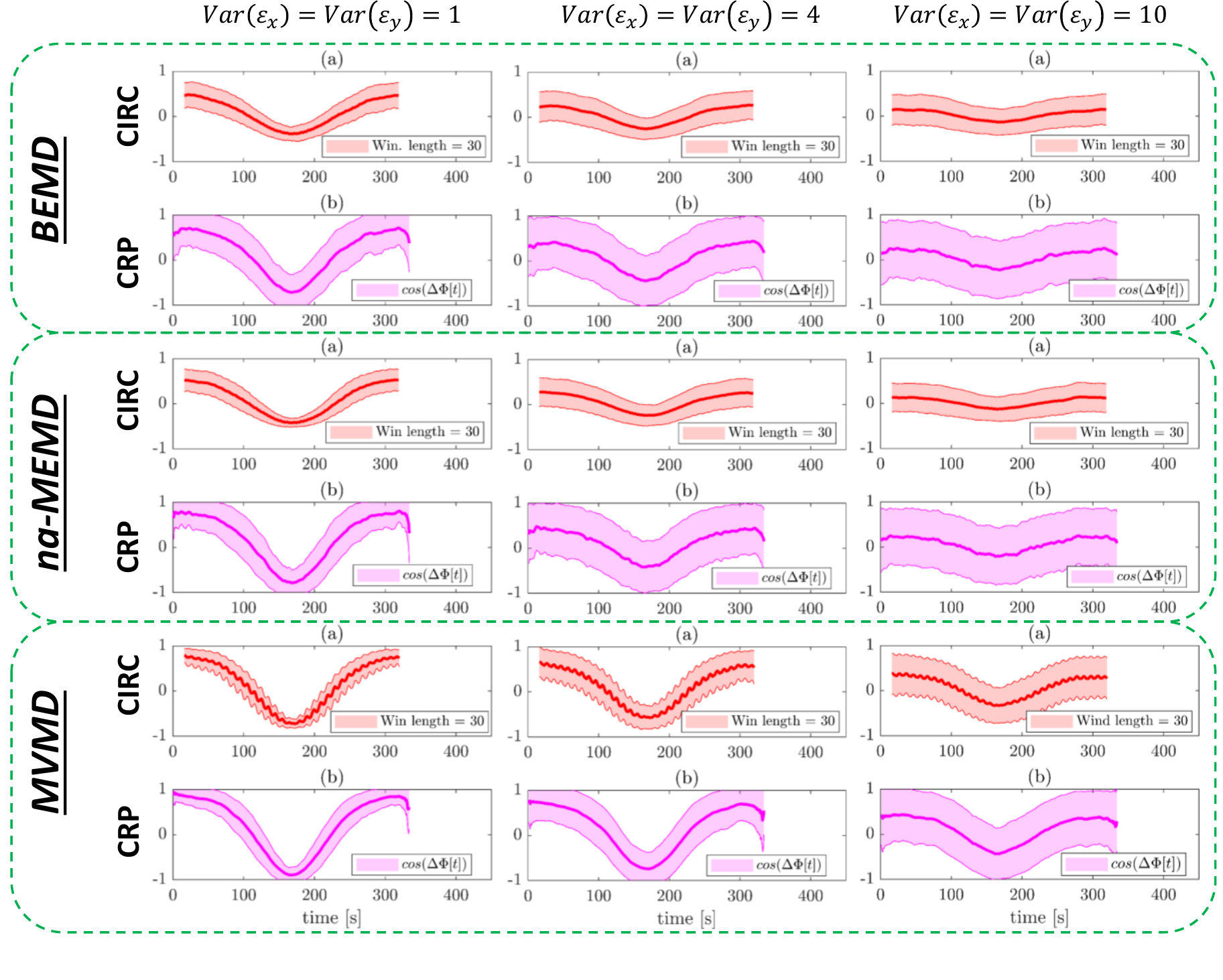}
\end{center}
\caption{Results of Simulation 6.  A comparison of BEMD, na-MEMD, and MVMD-based PS under three different noise levels (1, 4, and 10). The ground true phase shift is a sigmoid function as described in Simulation 3. Results are based on an IMF extracted with a central frequency of $0.05 \, Hz$. Results are shown for tWPS using circular-circular correlation (CIRC) and IPS using cosine of relative phase (CRP).  In each panel, the mean and 95\% interval for each measure are shown at each time point.  }\label{fig: PerformanceComparison_PSMeasures}
\end{figure}

Figures \ref{fig: Sim1_PSMeasures1000} - \ref{fig: PerformanceComparison_PSMeasures}, show the results from the six simulation studies. For each we compare three variants of MD, namely BEMD, na-MEMD, and MVMD.  Two measures of PS, tapered WPS using circular-circular correlation (CIRC) and IPS using cosine of relative phase (CRP)) were used to evaluate the time-varying PS for data obtained using each MD technique. Hence, there are six different combinations of MD techniques and PS measures evaluated in each simulation. Below follows discussion of the results of each simulation study in turn.

{\em Simulation 1:} 

Figure \ref{fig: Sim1_PSMeasures1000} shows the results for Simulation 1. The mean and 95\% confidence interval for each measure are shown at each time point for each combination of MD technique and PS measure.  For the tWPS measures results are shown for window lengths of 30 60, and 120 time points. The analysis was performed using the IMF centered at $0.05\, Hz$, as this corresponded to the central frequency of the simulated signal in Simulations 2 and 3.

The results indicate in the null case, for each MD technique both CIRC and CRP yield a mean time-varying PS measure of about 0.  CRP performs roughly equivalently across the three MD approaches. For CIRC, BEMD and na-MEMD perform similarly, while MVMD provides a narrower confidence interval.  In general, the confidence intervals obtained using CIRC are narrower than those obtained using CRP. In addition, the confidence intervals become increasingly narrow for CIRC as the window length increases. This is to be expected as the PS measure is constant across time and averaging over more time points should provide more stable estimates.

{\em Simulation 2:} 

Figure \ref{fig: Sim2_PSMeasures1000} shows the results for Simulation 2.  Each MD technique successfully decomposed the monocomponent signal pairs used in the simulation correctly, leading to one pair of IMFs for each approach with a central frequency of $0.05 \, Hz$.
Using these IMF pairs, the phase synchronization analysis was performed using both PS measures.   The mean and 95\% confidence interval for each measure are shown at each time point for each combination of MD technique and PS measure.

Since the generated signal pairs in this simulation consisted of a monocomponent sinusoidal signal with a phase shift corresponding to a ramp function (see Panel (a)), the signal will be in phase up to $t = 170\;s$ and afterwards the phase shift linearly increases with time allowing the signal pair transition to being out of phase (at $t = 210\;s$ and $290\;s$) and back into phase.  

Each combination of MD technique and PS measure was able to successfully capture the transition in and out of phase between the signal pairs.  Consistent with the null setting in Simulation 1, the performance of BEMD and na-MEMD are roughly equivalent. In contrast, MVMD leads to narrower confidence intervals and achieves better performance. In addition, CRP generally outperforms CIRC particularly when used together with MVMD. This can be seen by the fact that CRP is able to better capture when the signals are completely in or out of phase by taking values close to 1 and -1 in these situations. CIRC's ability to capture transitions in phase worsens as the window length increases. This is not surprising as longer window lengths make it more difficult to capture more transient changes in PS.

{\em Simulation 3:} 

Figure \ref{fig: BEMD_Sim3_PSMeasures1000} shows the results for Simulation 3.  Each MD technique successfully decomposed the monocomponent signal pairs used in the simulation correctly, leading to one pair of IMFs for each approach with a central frequency of $0.05 \, Hz$. 
Using the IMF pair obtained using each approach, the phase synchronization analysis was performed using both PS measures.   The mean and 95\% confidence interval for each measure are shown at each time point for each combination of MD technique and PS measure.

Since the generated signal pairs in this simulation consisted of a monocomponent sinusoidal signal with a phase shift corresponding to a sigmoid function (see Panel (a)), the signals will initially be in phase, after which the amount of phase shift gradually increases. This continues until $t=170$ when the pairs are in anti-phase synchronization. Thereafter, the signals gradually return to being in phase.  The transition between the phase of the signals from 0 to $2\pi$ occurs smoothly and monotonically increasing. 

Each combination of MD technique and PS measure was able to successfully capture the transition in and out of phase between the signal pairs. Consistent with the null setting in Simulation 1, the performance of the BEMD and na-MEMD perform roughly equivalently. In contrast, MVMD leads to a narrower confidence intervals and achieves better performance. In addition, CRP generally outperforms CIRC particularly for MVMD. 
Finally, CIRC's ability to capture transitions in phase improves as the window length increases. This is consistent with the slower transitions in and out of phase as compared with Simulation 2.

{\em Simulation 4:} 

Figure \ref{fig: Sim4_PSMeasures} shows the results of Simulation 4.  Each MD technique was used to obtain one pair of IMFs for each approach with a central frequency of $0.05 \, Hz$. Using these IMF pairs, the phase synchronization analysis was performed using both PS measures.   The mean and 95\% confidence interval for each measure are shown at each time point for each combination of MD technique and PS measure.

The results indicate that neither EMD approach (i.e., BEMD or na-MEMD) is able to successfully separate the frequency components from a multicomponent signal when the frequencies are close to each other. This can, for example, be seen in panels (b) and (c), which show that PS analysis based on the use of BEMD does not accurately reflect the ground truth phase relationship between signals.  The issue is that BEMD does not adequately address mode mixing, which is a well known shortcoming of EMD-based approaches in this setting.  Mode mixing leads to the appearance of both frequency components in the signal in the extracted mode obtained by BEMD.  This manifests itself as an amplitude modulation in the extracted mode and consequently the phase estimates, leading to the additional humps seen in the estimated PS measures.  Similar results are obtained when using na-MEMD.  While the introduction of noise is meant to to mitigate the mode mixing problem and untangle closely located spectra, it does not appear to have worked in this simulation. While panels (d) and (e) show tighter confidence intervals and  smoother estimates of PS compared to BEMD (panels (b) and (c)), the effects of amplitude modulation is still present.
In contrast, the results shown in panels (f) and (g) indicate that MVMD was able to successfully separate the frequencies.  When performing PS analysis based on the use of MVMD, the extracted mode accurately retrieved the component of interest from the bivariate signal.  The results show that the PS measures based on the extracted mode are able to accurately capture the extent of phase synchronization at each time point.  

Both estimated PS measures (CIRC and CRP) are able to capture the ground truth phase shifts. This can be seen as both measures are close to 1 at the beginning of the signal. As the signals  transition to an anti-phase state at $175\,[s]$, the PS measures approach -1 and as the signals come back into phase, the measures approach 1. CRP generally outperforms CIRC as it is able to better capture when the signals are completely in or out of phase by taking values close to 1 and -1 in these situations.

{\em Simulation 5:} 

Figure \ref{fig: multivariate_Sim4_PSMeasures} shows the results of Simulation 5. For each MD technique panel (a) shows the ground truth time intervals when the signals are in each of the three states; panel (b) shows a heat map depicting when state transitions occur for both PS measures across the 1000 realizations of the simulation; and panel (c) shows the three estimated states averaged across the 1000 realizations.  The results are shown both for CIRC with window length 30 and CRP.  

Studying the results for BEMD, the estimated states appear to capture the relationship between the signals within each epoch.  For State 1, the ground truth states that $x_1(t)$ and $x_3(t)$ are in anti-phase with $x_2$, while in phase with one another.  The estimated State 1 shows that $x_1(t)$ and $x_2(t)$ are negatively phase synchronized (-0.4 for CIRC, and -0.6 for CRP), $x_1(t)$ and $x_3(t)$ are positively phase synchronized (0.45 for CIRC and 0.62 for CRP), and $x_2(t)$ and $x_3(t)$ are negatively phase synchronized (-0.4 for CIRC, and -0.6 for CRP).  The estimated State 1 corresponds to ground truth State 1 shown in panel (a) in the time interval $150-250\,[s]$.  While the BEMD-based analysis captures the pairwise relationship between signals, the PS estimates are lower than expected, as according to the ground truth  $x_1(t)$ and $x_3(t)$ should be completely in phase. Further the estimated values using CIRC are lower than those obtained using CRP, which is related to the additional smoothing associated with using a windowed approach. These results are consistent with the previous simulations. For State 2, the ground truth states that all signals should be in phase. The estimated State 2 largely captures this behavior.  However, the estimates are again lower than the ground truth values of 1 (0.45 for CIRC and 0.6 for CRP). For State 3, the ground truth states that $x_1(t)$ and $x_2(t)$ are in anti-phase with $x_3(t)$, while in phase with one another. The estimated State 3 shows that $x_1(t)$ and $x_2(t)$ are positively synchronized, while $x_3(t)$ is negatively associated with $x_1(t)$ and $x_2(t)$. The estimated CRP value between $x_1(t)$ and $x_3(t)$ is -0.55 and between  $x_2(t)$ and $x_3(t)$ is -0.35.  Note the difference in the estimated values can be attributed to the fact that BEMD performs mode alignment in a pairwise manner, which can manifest itself in misalignment of the modes across pairs and discrepancies in the estimated PS measures. Finally, the heat map of state transitions shown in panel (b) demonstrates the state switching behavior for the 1000 realizations in the simulation. On average, the method is able to capture the transitions from one state to another. However, there remain significant missclassification of states and abrupt switching between states is observed.

The results for na-MEMD show that the differences between PS estimates due to mode misalignment that appeared in the BEMD-based analysis have been addressed by the multivariate nature of the  mode extraction performed by na-MEMD.  Further as seen in panel (c), the estimated values are closer to the ground truth values.  For example, in State 1 $x_1(t)$ and $x_3(t)$ are negatively associated with $x_2(t)$ ($-0.4$ for CIRC and $-0.8$ for CRP).  This improved performance can be attributed to the properties of na-MEMD, including mitigation of mode-mixing and the joint extraction of the mode corresponding to central frequency of the generated trivariate signal. This improvement can be similarly seen in the other states; see panel (c). Further the transition between states shown in panel (b) demonstrates improved state classification.  In addition, the  transition times at $t = 50, 125, 150, 250\,[s]$ are better captured by CRP in combination with na-MEMD.  

Finally, due to its multivariate nature and reduced sensitivity to noise, the MVMD-based analysis is able to address the mode mixing issues and obtain values close to the ground truth. For example, in State 2 the signals are almost all in complete positive phase synchrony with one another. The signals $x_1(t)$ and $x_2(t)$ are positively phase synchronized (0.72 for CIRC and 0.85 for CRP), as are the signals $x_1(t)$ and $x_3(t)$ (0.75 for CIRC and 0.91 for CRP).  These results carry over to the other estimated states, resulting in closer estimates of the ground truth PS values using MVMD. In addition, as seen in panel (b), the misclassification of state transitions are less pronounced compared to the other MD approaches.

{\em Simulation 6:} 

Figure \ref{fig: PerformanceComparison_PSMeasures} shows the results for Simulation 6.  Each MD technique successfully decomposed the monocomponent signal pairs used in the simulation correctly, leading to one pair of IMFs for each approach with a central frequency of $0.05 \, Hz$. Using the IMF pair obtained using each approach, the phase synchronization analysis was performed using both PS measures. The mean and 95\% confidence interval for each measure are shown at each time point for each combination of MD technique and PS measure. 

The simulations are equivalent to Simulation 3, with a phase shift corresponding to a sigmoid function. However, the results are evaluated at different noise levels. For a fixed level of noise, the estimates of the phase synchronization for MVMD are closer to the ground truth compared to the two other approaches.  This is particularly pronounced looking at $t=170\,[s]$ where the two signals created are in complete anti-phase.  The average estimated CRP value is -0.78 using BEMD, -0.81 using na-MEMD,  and -0.92 using MVMD.  As expected, the performance worsens as the noise level increases. This can be seen by the widening of the confidence intervals and the reduced ability to detect when the signal is in complete phase or anti-phase. MVMD performs better for lower noise levels, but the MD approaches are roughly equivalent as the noise becomes larger.

\newpage \subsection{Analysis of HCP Data}

\begin{figure}[!ht]
\begin{center}
\includegraphics[width=0.8\textwidth,trim={1.3cm 1cm 1.5cm 2cm},clip]{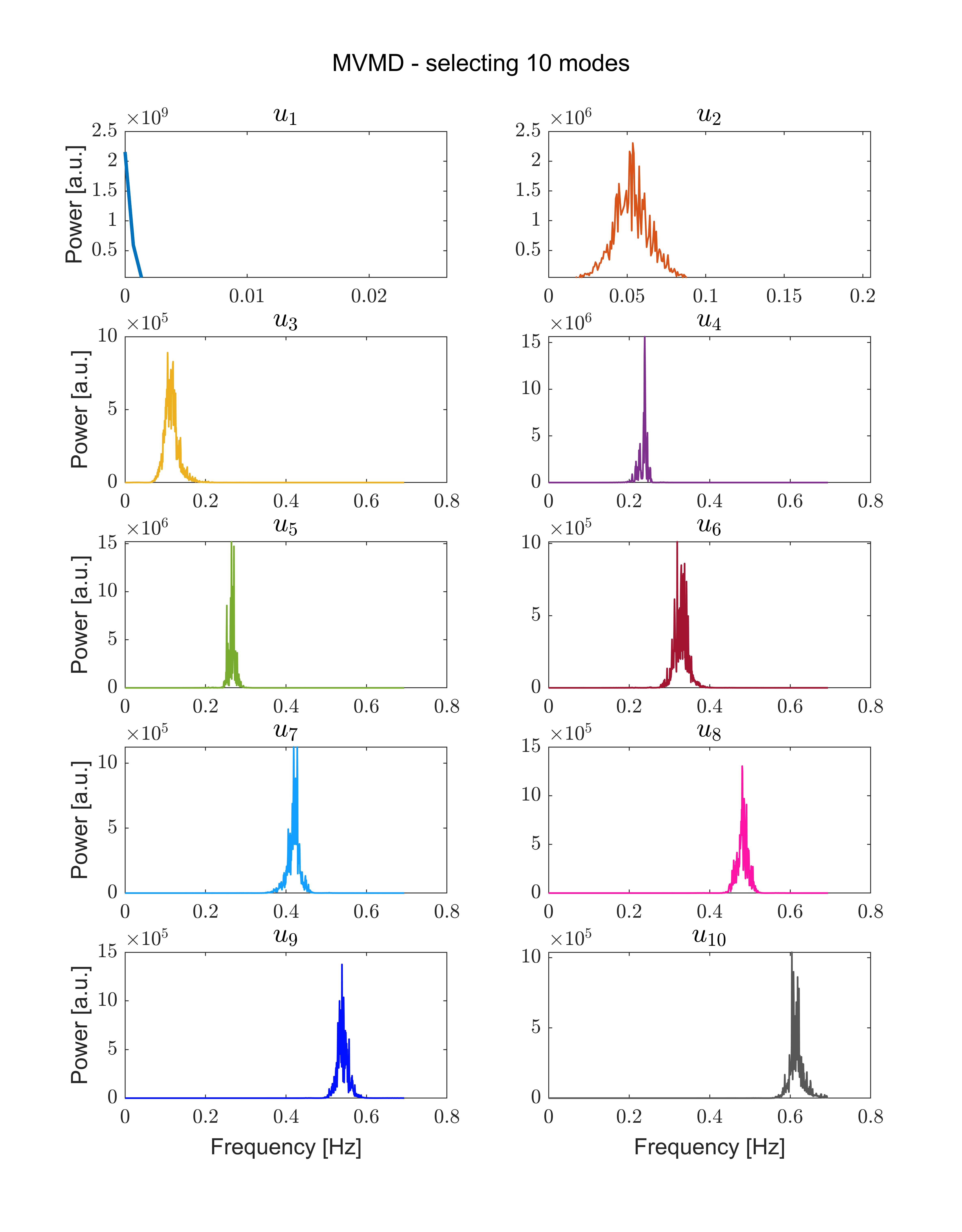}
\end{center}
\caption{Analysis of the HCP Data. The power spectral density summed over all subjects for the first ten IMFs obtained by applying MVMD to the 101 signals (100 brain nodes $+$ one respiration node). }\label{fig: PSDMVMD}
\end{figure}

Figure \ref{fig: PSDMVMD} shows the power spectral density (PSD) summed over all subjects for the ten IMFs obtained by applying MVMD to the 101 signals. Mode $u_1$ has its peak close to $0.001\,Hz$ and corresponds to signal drift commonly seen in fMRI data. Mode $u_2$ has the highest PSD in the range $0.01-0.1\,Hz$, which is the standard  frequency band used in rs-fMRI studies. Mode $u_4$ has the highest PSD in the range close to the respiration frequency. We focus further analysis on modes $u_2$ and $u_4$ as the former corresponds to a more standard analysis of rs-fMRI data and the latter allows us to determine which brain regions are phase synchronized with the respiratory node. After applying CRP to the extracted multivariate signals from each mode, the time-varying connectivity measures were clustered into 2 reoccurring brain states. 

\begin{figure}[!ht]
\begin{center}
\includegraphics[width=\textwidth,trim={0cm 1cm 0.0cm 2cm},clip]{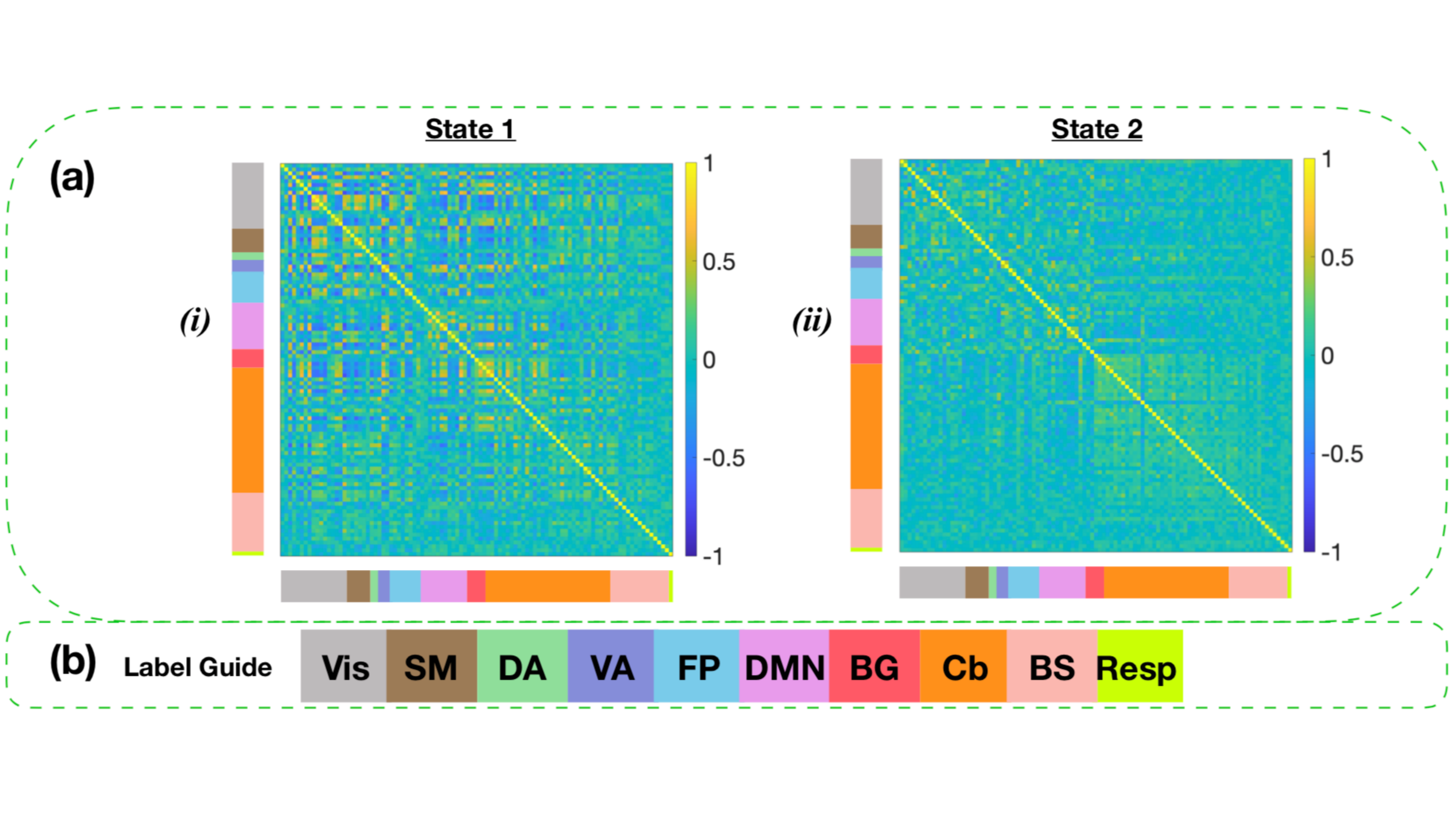}
\end{center}
\caption{Analysis of the HCP Data. In panel (a), (i) \& (ii) show the two reoccurring brain states of PS corresponding to mode $u_2$ (0.045 Hz) obtained using MVMD-based PS using CRP on 101 nodes where node 101 corresponds to the respiratory node.  Panel (b) provides labels for the various  regions.  
}\label{fig: HCPtvps2state_0.045Hz}
\end{figure}

Figure \ref{fig: HCPtvps2state_0.045Hz} shows the two brain states corresponding to mode $u_2$ which has its peak PSD at a central frequency of $0.045\,Hz$. The states are organized into nine networks (visual, somatomotor, dorsal attention, ventral attention, frontoparietal, default mode network (DMN), basial ganglia, cerebellum, and brainstem).  State 1 shows moderate to high correlations among signal components representing somatomotor, default mode, and dorsal attention networks.  In addition, a set of components in the cerebellum show negative correlations with visual, somatomotor, frontoparietal, default mode, and attention networks. State 2 shows moderate to high correlations within the cerebellum and within the visual, somatomotor, frontoparietal, default mode, and attention networks. 

\begin{figure}[!ht]
\begin{center}
\includegraphics[width=0.93\textwidth,trim={6.2cm 1.5cm 6cm 0cm},clip]{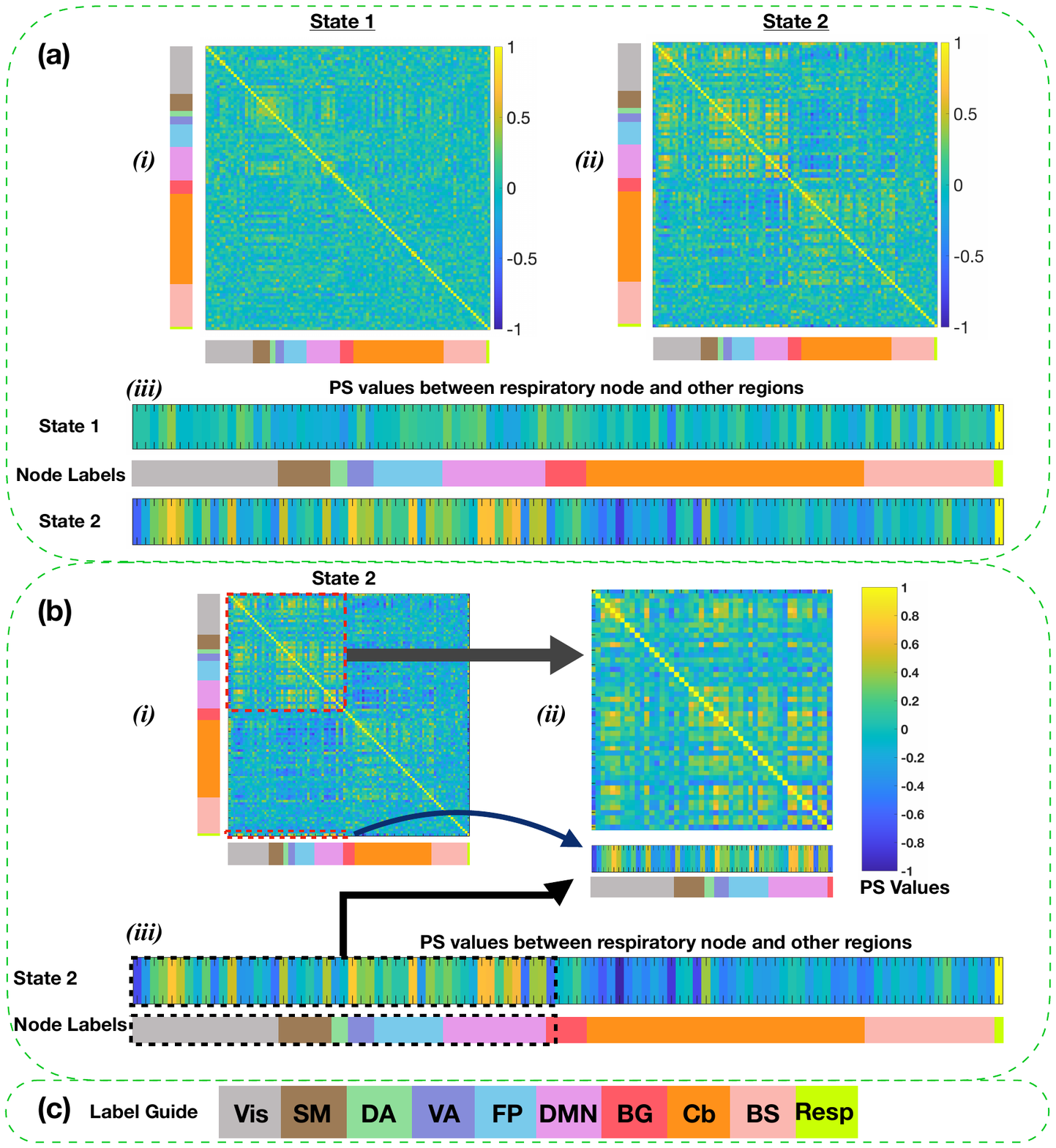}
\end{center}
\caption{Analysis of the HCP Data. In panel (a), (i) \& (ii) show the two reoccurring brain states of PS corresponding to mode $u_4$ (0.27 Hz) obtained using MVMD-based PS using CRP. The bottom portion (a.iii) shows PS between the respiratory node and the other 100 rs-fMRI brain nodes for States 1 and 2.  Here node 101 corresponds to the respiratory node. Panel (b.i) shows State 2, with the nodes hyperconnected with respiratory signal marked with a dashed red line.  Panel (b.ii) shows a close-up look at regions with high degree of PS with the respiratory node. These include regions in the visual, somatomotor, dorsal attention, ventral attention, frontoparietal, and default mode networks. Panel (b.iii) provides a closer look at the PS values between respiratory node and various regions.  Panel (c) provides labels for the various  regions.}\label{fig: HCPtvps2states}
\end{figure}

Figure \ref{fig: HCPtvps2states} shows the two brain states corresponding to mode $u_4$ which has its peak PSD at a central frequency of $0.27\,Hz$. The states are organized into the same nine networks as above plus the respiration node. The top portion of panel (a) show the two reoccurring brain states, while the bottom portion shows the PS between the respiratory node and the 100 rs-fMRI brain nodes for each of States 1 and 2.  State 1 shows moderate to high correlations among signal components representing somatomotor, default mode, and attention networks.  In State 2 a set of components in the cerebellum show negative correlations with visual, somatomotor, frontoparietal, default mode, and attention networks. These negative correlations were not observed in State 1.  In State 1 the respiratory node does not appear to play a significant role in PS with the brain, except at a single node in the cerebellum which is negatively associated with the respiratory signal.  However, in State 2, there are several nodes that are significantly phase synchronized with respiration. For example, several nodes from the visual, somatomotor, ventral attention, frontoparietal, and DMN are positively associated with respiration, while nodes in the cerebellum are negatively associated with respiration.  Interestingly, the nodes that are highly phase synchronized with the respiratory signal occur within regions that are themselves hyper-connected in this state.  To illustrate, in panel (b) we highlight networks that show a positive phase synchronization with the respiratory signal.  Clearly, these networks also show strong within-network synchronization.

\section{Discussion}

There is growing interest in measuring time-varying functional connectivity between time courses from different brain regions using resting-state fMRI (rs-fMRI) data. One approach towards achieving this goal is to measure the phase synchronization (PS) between regions across time. However, this type of analysis requires that the data be bandpass filtered prior to computing PS measures. This, in turn, requires the \textit{a priori} choice of both the type and cut-off frequencies for the filter.  

The need to bandpass filter the data stems from the fact that for the instantaneous phases obtained using the Hilbert transform to be physically meaningful, the signal must be sufficiently narrow-bandpassed to approximate a monocomponent signal. As an alternative, researchers are increasingly exploring the use of various mode decomposition (MD) techniques that allow for the data driven decomposition of signals into narrow-band components (referred to as intrinsic mode functions (IMFs)) centered at different frequencies, thereby providing a substitute for bandpass filtering. In this paper we have  explored several variants of MD, including empirical mode decomposition (EMD), bivariate EMD (BEMD), noise-assisted multivariate EMD (na-MEMD), and multivariate variational mode decomposition (MVMD) in the context of estimating time-varying PS. 

When working with rs-fMRI data, it is generally of interest to extract IMFs from a collection of signals from multiple regions of interest (ROIs). One approach is to apply the standard EMD algorithm separately to each signal separately and extract a set of IMFs from each. However, this will not guarantee that the frequencies of the extracted IMFs match across signals, nor will the repeated application of the EMD algorithm necessarily produce the same number of IMFs for all signals. These issues complicate subsequent analysis of PS across regions. To circumvent them and allow for the decomposition of a bivariate or multivariate signal, various extensions of EMD have been proposed that perform the decomposition jointly \cite{ur2011filter}.  For example, BEMD performs the decomposition on each pair of signals separately. However, if analyzing more than two signals this leads to similar problems described above. In this setting it is instead beneficial to use multivariate approaches such as na-MEMD and MVMD that perform decomposition jointly on all signals. 
 
While EMD-type decomposition's provide a powerful tool, they can suffer from a mode-mixing phenomena when analyzing real signals.  This refers to a situation where an IMF consists of contributions from multiple frequencies \cite{xu2019causes}, and occurs when oscillations with different time scales coexist in the same IMF or when  oscillations in the same time scale are assigned to different IMFs. To circumvent this problem, noise-assisted techniques such as na-MEMD introduce noise signals that alleviate the mode-mixing problem.  Thus, in settings where there is the potential for any mode-mixing of the input signals, which is often a case in the analysis of rs-fMRI data, the use of na-MEMD can help mitigate this problem to some extent.

In addition to issues such as mode-alignment and mode mixing (which na-MEMD addresses to some extent), the class of EMD techniques are influenced by a sampling induced deviation. This is because they require the computation of signal extrema, whose value can be easily influenced by how finely a signal is sampled.  A consequence is that the local mean may introduce artefacts related to the manner in which sampling is performed \cite{rilling2009sampling}. Furthermore, EMD-type approaches tend to be susceptible to noise. Variational mode decomposition (VMD), and its multivariate version MVMD, have shown a robustness to noise in the input signal due to the inclusion of the parameter $\tau$ in Equation \ref{eqn: LagrangianMultip} that allows one to control and regulate Lagrangian multipliers. When the input signal is noisy, decomposing signal that fully reconstructs the input signal might not be desirable.  By setting the aforementioned parameter to zero, the robustness can be controlled within VMD and
MVMD. In contrast, in EMD approaches there is no way of hindering noise from entering into the decomposition process \cite{dragomiretskiy2013variational, ur2019multivariate}.

In addition to comparing MD techniques, in this work we also investigate how they combine with various approaches towards measuring PS. This includes methods that combine a PS metric with a sliding window approach, referred to as windowed phase synchronization (WPS), with those that directly measure PS at each time point, referred to as instantaneous phase synchronization (IPS) \cite{honari2021evaluating}.  Within the WPS framework we use circular-circular correlation (CIRC) to assess PS, while within the IPS framework we use cosine of relative phase (CRP). In addition, in this work we used a tapered sliding window approach in contrast to earlier work where we used a rectangular window \cite{honari2021evaluating}. In general, we find that WPS estimates are more accurate when using a tapered window \cite{honari2021measuring}, as the use of a tapered window allows for smoother results and avoids artifacts that arise due to the sudden changes related to the edges of a boxcar window \cite{lindquist2014evaluating, shakil2017parametric}.   

The six simulations illustrate several important points regarding the performance of these methods. For each simulation we combined three different MD techniques (BEMD, na-MEMD, and MVMD) with the Hilbert transform to compute the phase. Next, time-varying PS measures were computed using tapered WPS in combination with CIRC and IPS evaluated using CRP \cite{honari2021evaluating}. Thus, in total six different combinations of MD techniques and PS measures were assessed. 

In Simulations 1-3, the performance of BEMD and na-MEMD are roughly equivalent. In contrast, MVMD led to a narrower confidence interval and achieved better performance. The similarity in performance between BEMD and na-MEMD is not surprising as the signal being analyzed is bivariate and the benefits of using a general multivariate approach are not yet apparent. When analyzing the null data (Simulation 1) the confidence interval obtained using CIRC were consistently narrower than those obtained using CRP. However, when true time-varying PS exists (Simulations 2-3), CRP generally outperformed CIRC, particularly when used together with MVMD. This can be seen by the fact that CRP was better able to capture when the signals are completely in or out of phase by taking values close to 1 (in-phase) and -1 (out of phase) in these situations.

In Simulation 4, the ability of the MD techniques to handle mode mixing were assessed.  Here mode mixing led to the appearance of two frequency components in the extracted mode obtained using BEMD. This, in turn, led to an amplitude modulation in the extracted mode causing the additional humps seen in the estimated PS measures; see Figure \ref{fig: Sim4_PSMeasures}. Similar results were obtained using na-MEMD. While the introduction of noise is meant to mitigate the mode mixing problem and to some degree untangle closely located spectra, it does not appear to have worked in this simulation.
In contrast, MVMD was able to separate the frequency components and avoid amplitude modulation and mode mixing, providing superior capability to assess PS compared to both BEMD and na-MEMD.

In Simulation 5, the MD techniques were compared when working with a multivariate signal.  In this setting multivariate MD approaches (e.g., na-MEMD and MVMD) outperformed BEMD.  This is expected due to the fact that the multivariate approaches are able to address the issue of mode misalignment during the decomposition of multivariate signals.   The simulation also assessed the performance of the combined MD techniques and PS measures when estimating PS states and transitions between states.  Here MVMD outperformed the other MD techniques in terms of accurately estimating the pairwise PS values that make up each state and capturing the timing of transitions between states.

In Simulation 6, we evaluated the ability of the different combinations of MD techniques and PS measures to overcome increasing noise levels.  This is particularly important in settings where the SNR is low, as is the case in the analysis of fMRI signals.  The performance of the MD techniques were evaluated at three different noise levels, and it was shown that MVMD outperformed the other approaches by obtaining estimates closer to the ground truth. Additionally, it provided narrower confidence interval for the estimates.  However, as the noise level increased, the performance of the different MD techniques become roughly equivalent, indicating that in very low SNR settings none of the methods may work very well.

In the simulations, the effect of the window length for the tWPS measures were also investigated.  The results indicated that longer window lengths tend to provide more accurate estimates of PS as they led to a decrease in the variation of the estimates. However, using longer windows made it harder to detect subtle changes. Thus, longer window lengths are better for detecting slower phase shifts (see Simulations 1 and 3), while shorter window lengths are better at capturing more rapid changes (see Simulation 2).

The results of our simulations indicate that MVMD together with CRP provided superior performance compared to the other combinations of MD techniques and PS methods. Therefore, the proposed MVMD-based IPS framework was applied to real rs-fMRI data on 50 subjects from the Human Connectome Project (HCP).  Here, a physiological node corresponding to a respiration time series, was also included.  The multivariate signal, corresponding to 100 brain nodes and the respiration node, were decomposed using MVMD.  The results were used to investigate not only the relationship between functional brain networks, but also assess the PS of functional networks and the respiratory node.  After studying the power spectrum of the modes extracted by MVMD, two of them were chosen for PS analysis.  The first mode, which peaks at a central frequency of $0.045\,Hz$, was chosen because it is located in the frequency band $0.01-0.1\,Hz$ commonly analyzed in rs-fMRI studies.  The second mode is chosen based on the fact that its central frequency is close to the respiration frequency.  Performing PS analysis on this mode shows that several rs-fMRI networks are highly phase synchronized with the respiratory node in one of the estimated brain states.  The nodes that are highly synchronized with respiration also showed a higher degree of within-network PS.  Interestingly, the PS relationship within these networks closely resemble the strength of PS between the respiratory node and the networks.

In summary, in this work we have compared a number of MD techniques using both simulations and rs-fMRI data in the context of assessing time-varying PS.  Our results show that MVMD outperforms other evaluated MD approaches, and further suggests that this approach in combination with CRP can be used to reliably investigate time-varying PS in rs-fMRI data.

\section*{Acknowledgments}
The work presented in this paper was supported in part by NIH grants R01 EB016061 and R01 EB026549 from the National Institute of Biomedical Imaging and Bioengineering and R21 NS104644 from the National Institute of Neurological Disorders and Stroke.

\newpage
\bibliographystyle{model2-names}

\end{document}